# THE METALLICITIES OF STAR-FORMING GALAXIES AT INTERMEDIATE REDSHIFTS $0.47 < z < 0.92$

SIMON J. LILLY [1,2], C. MARCELLA CAROLLO[1,2]
AND
ALAN N. STOCKTON[3,4]

## ABSTRACT

Estimates of the [O/H] metallicity of the star-forming gas in a sample of 66 CFRS galaxies between $0.47 < z < 0.92$ have been made based on the flux ratios of bright emission lines. The spectra of the galaxies are very similar to those of typical galaxies in the local Universe. Most galaxies (> 75%) have the [O/H] ~ 8.9 metallicities that are seen locally in galaxies of similar luminosities. However, a minority (< 25%) appear to have significantly lower metallicities [O/H] < 8.6 as indicated by high values of the Pagel $R_{23}$ parameter. The fraction of the sample at these lower [O/H] would be reduced if the reddening in these objects was lower than the $E_{B-V}$ ~ 0.3 assumed. The high metallicities of the majority of the galaxies suggest that they do not fade to be low metallicity dwarf galaxies today. Only one of the 66 galaxies has an optical emission line spectrum similar to the few Lyman break galaxies recently observed at $z$ ~ 3, emphasizing the differences with that population. The inferred emission line gas [O/H] metallicity broadly correlates with luminosity in both the rest *B*- and *J*-bands but with considerable scatter introduced, especially at lower luminosities, by the range in [O/H] observed. The metallicity does not appear to correlate well with galaxy size, Hβ strength or, with the very limited data available, the kinematics. The metallicity does correlate well with the continuum optical-infrared colors in a way that could be explained as a combination of photospheric effects, differing ages of the stellar population and/or different amounts of reddening. None of these should produce large variations in the M/L ratio. These results support a "down-sizing" picture of galaxy evolution in which the manifestations of star-forming evolutionary activity appear in progressively more massive galaxies at earlier epochs rather than a "fading dwarf" picture in which the luminous active galaxies at high redshift are highly brightened dwarf galaxies. The overall change in metallicity of star-forming galaxies over the last half of the age of the Universe appears to have been modest, Δ[O/H]= 0.08 ± 0.06. This is consistent with the age-metallicity relation in the Galactic disk and is broadly consistent with models for the chemical evolution of the Universe, especially those that consider different environments.

[1] Department of Physics, Swiss Federal Institute of Technology (ETH-Zurich), ETH Hˆnggerberg, CH-8093 Zurich, Switzerland
[2] Visiting Astronomer, Canada-France-Hawaii Telescope, jointly operated by the National research Council of Canada, the Centre National de la Recherche Scientifique of France and the University of Hawaii.
[3] Institute for Astronomy, University of Hawaii, 2680 Woodlawn Drive, Honolulu, Hawaii 96822, U.S.A.
[4] Visiting Astronomer, W.M. Keck Observatory, jointly operated by the California Institute of Technology and the University of California.

## 1. INTRODUCTION

The metallicity of the Universe and of objects in it provides a fundamental metric reflecting the development of structure and complexity in the Universe on galactic scales. This metric is all the more important because it is relatively easily observable and "long-lived" in the sense that heavy atomic nuclei, once produced, are not readily destroyed.

For many years, the metallicities of gas at high redshifts have been studied through the analysis of absorption line systems seen in quasar spectra. The lines of sight to quasars probe, almost by definition, random regions of the Universe - the only possible concern being whether lines of sight passing through dusty regions are under-represented (Fall and Pei 1993, see Ellison et al. 2001). On the other hand, only a single line of sight through a given system is generally available, making the interpretation of the metallicity measurement in the context of larger structures, such as galaxies, non-trivial and fundamentally statistical in nature. As an example, the relationship between the high column density "damped Lyman $\alpha$" (DLA) systems and galaxies is still by no means clear.

The study of the metallicities of material in known galaxies at high redshift is at a much earlier stage of development and is inevitably of lower sophistication.

However, metallicity estimates, especially of the [O/H] abundance, using diagnostics that are based solely on strong emission lines, e.g., [OII] $\lambda3727$, H$\beta$, [OIII] $\lambda\lambda4959,5007$, H$\alpha$, [NII] $\lambda6583$, [SII] $\lambda\lambda6717,6731$, are now technically feasible over a wide range of redshift and give a new perspective on the chemical evolution of galaxies.

For some purposes, the estimates of the metallicity of star-forming gas at earlier epochs that are obtained from emission lines may actually be of more relevance than the more "global" view that is obtained from the absorption line studies. Applications include the confrontation of models for the chemical evolution of the Milky Way or of other galaxies and the use of metallicity estimates to constrain the present-day descendents of high redshift galaxies. The $R_{23}$ parameter introduced by Pagel et al. (1979) provides an estimator for the [O/H] abundance.

$$(1) \quad R_{23} = \frac{f_{[OII]3727} + f_{[OIII]4959,5007}}{f_{H\beta}}$$

$R_{23}$ has some undoubted drawbacks: It is sensitive to reddening and has two metallicity solutions for most values of $R_{23}$. It also depends on ionization, though this can be determined relatively easily from the oxygen lines. These issues have been amply discussed in the literature (see e.g., Kobulnicky, Kennicutt & Pizagno 1999, Kewley & Dopita 2002). The great virtue of $R_{23}$ is that it is based on a few strong emission lines and can therefore be applied in a uniform way over a wide range of redshifts. Some of the difficulties in calibration can presumably be circumvented by considering relative effects rather than relying on absolute determinations of metallicity. The aim of this paper is to search for and interpret differential effects between different redshifts rather than to determine absolute metallicities for individual galaxies.

We have been undertaking a program of spectroscopy of star-forming galaxies selected from the Canada-France-Redshift Survey (CFRS; Le Fevre et al 1995, Lilly et al. 1995, Hammer et al 1995) at intermediate redshifts $0.5 < z < 1.0$. The first results of this program (based on a sample of 15 galaxies) were reported in Carollo & Lilly (2001, hereafter Paper 1). We have now observed 94 galaxies in total from which we can form a statistically complete sample of 66 galaxies. This Paper presents our interpretation of these data. Other relevant studies that have been published recently include those of Kobulnicky & Zaritsky (1999) with 14 galaxies at $0.1 < z < 0.5$, Hammer et al. (2001) with 14 CFRS galaxies $0.45 < z < 0.8$ and Contini et al. (2002) with 68 mostly local, ultraviolet-selected, galaxies.

A "concordance" cosmological model with $H_0 = 70$ km s$^{-1}$ Mpc$^{-1}$, $\Omega_M = 0.3$ and $\Omega_{vac} = 0.7$ has been adopted throughout this paper. It should be noted that the effects of changing to an alternative model with $H_0 = 50$ km s$^{-1}$ Mpc$^{-1}$, $\Omega_M = 1$ and $\Omega_{vac} = 0$ is small at high redshift but would change the properties of the local comparison samples through the value of $H_0$.

## 2. OBSERVATIONS

### 2.1 *New $0.5 < \lambda < 1.0$ $\mu m$ spectroscopy at the CFHT*

In addition to the observations reported in Paper 1, new spectroscopy of CFRS galaxies with $0.45 < z < 1.0$ in six multi-object masks were obtained with the MOS spectrograph on CFHT during two further observing runs 25-27 February 2001 and 19-22 October 2001. The observational setup was similar to that in Paper 1 with a 1.3 arcsec wide slit giving a spectral resolution $R \sim 500$. Two significant improvements over our earlier data were adopted. First, a new filter blocking light shortward of 5500 Å was used, eliminating the possibility of second-order contamination of the spectra at all wavelengths below the CCD cutoff, albeit at the cost of a reduced spectral range at the blue end. Second, a 2k×4.5k 13.5 $\mu$m pixel EEV1 CCD was used, producing a

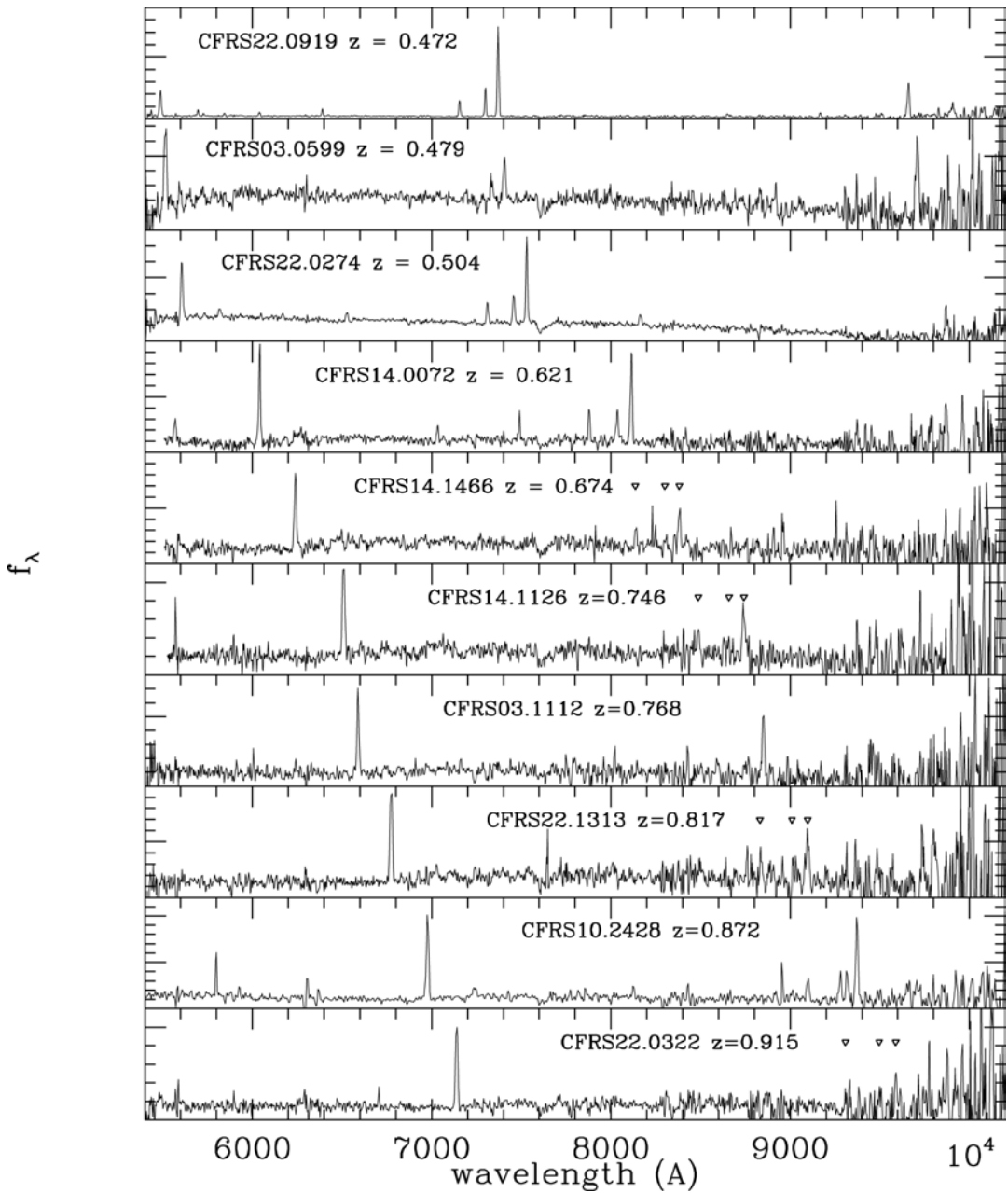

*Fig. 1 (a) The spectra of the ten galaxies in the final complete sample that exhibit the highest values of $R_{23}$, i.e. with metallicities near to the turn-around region, arranged in order of increasing redshift. Together with Fig 1b, these spectra illustrate the diversity seen in the CFRS sample at $0.5 < z < 1$. It should be noted that no attempt has been made to remove the atmospheric absorption feature at 7600 Å or to otherwise clean the spectra. Except where obvious, the locations of H$\beta$, and [OIII] 4959, 5007 are indicated.*

greatly improved sensitivity at the longer wavelengths. For each mask, nine independent 2700 sec exposures were obtained with the object at a different location along the 20 arcsec long slit. Subsequent processing of the data was effective in removing the prominent fringe pattern that is produced by this detector. The new spectra were photometrically calibrated using a total of 17 observations of the standard star Hiltner 600 obtained through a number of different slits in different multi-object masks. These displayed a consistency of better than 2% in the relative flux calibration at wavelengths $\lambda < 9200$ Å (deteriorating to about 10% at $\lambda > 9200$ Å) and a scatter in the absolute flux calibration of about 0.15 mag, mostly due to slit centering inaccuracies.

Representative spectra are shown in Figs 1a and 1b. Fig 1a shows the ten galaxies with highest $R_{23}$ while Fig 1b shows the ten galaxies with the lowest $R_{23}$. The range of line strengths in these figures illustrates the diversity of spectra in the population.

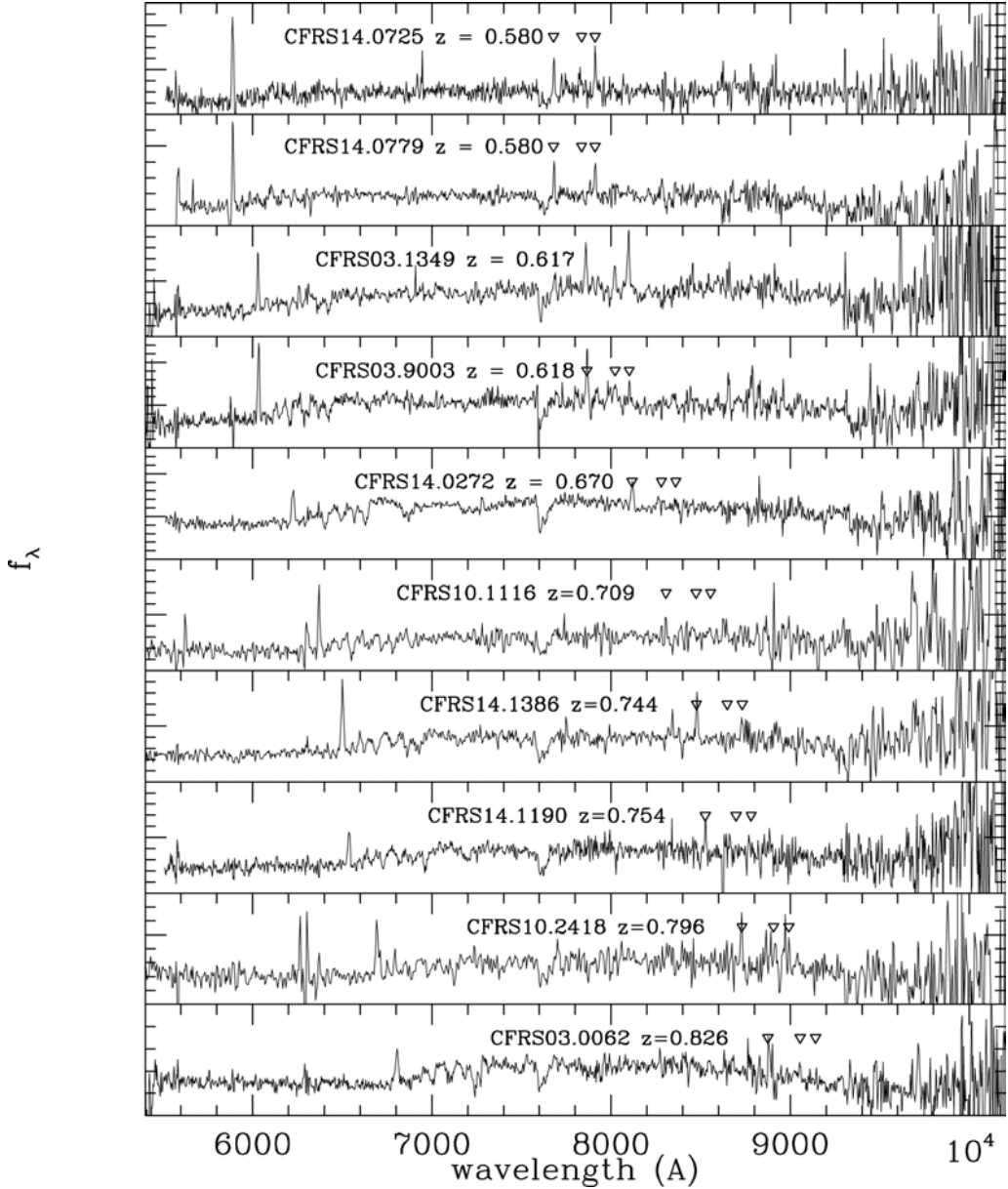

*Fig 1 (b) As in (a) except that the ten galaxies are those with the lowest values of $R_{23}$.*

Repeat observations of 14 of the 94 galaxies were made in more than one mask and, in most cases, during different observing runs. These repeat observations were reduced independently and show good agreement between the spectrophotometric measurements.

## 2.2 *J-band spectroscopy at Keck*

Four of the galaxies were observed with the NIRSPEC spectrograph on the 10-m Keck telescope in the spectral region contain H$\alpha$, and [NII] $\lambda$6583. One galaxy (CFRS14.0393) was observed during Keck observing time allocated through the Gemini Observatory and this observation was kindly carried out by Tom Geballe and Mirianne Takamiya. The remaining Keck observations were carried out on 2001 Jan 18 (UT). The instrumental setup used an 0.76 arcsec slit giving a spectral resolution of about 1200. In the present work, relative spectrophotometric calibration between the infrared and optical spectroscopy has not been attempted.

## 3. SAMPLE DEFINITION AND METALLICITY ESTIMATES

Emission line fluxes and equivalent widths were measured from the spectra using standard IRAF routines. In most cases, the uncertainty in the line flux is dominated by the uncertainty in the correct placement of the continuum and the quoted error bars are an attempt to reflect the possible limits. As in our earlier work, the effect of the underlying stellar Hβ absorption feature was accounted for by applying a correction to the rest-frame equivalent width of the Hβ emission line of +3 ± 2 Å. No other corrections have been applied to the line fluxes. Generally, the [OIII] λ4959 line is of poorer signal-to-noise ratio than [OIII] λ5007, and sometimes this line was badly affected by sky-subtraction or other problems. Since these lines have a ratio of 1:3 (in photon units) set by atomic parameters, it was decided to uniformly apply $f_{4959} = 0.34 f_{5007}$.

## 3.1 Sample definition

In any study of this nature, it is advantageous to have a statistically complete sample bounded by well-defined selection criteria. We have approached this in the following way, as illustrated in Fig. 2.

In populating the multi-slit masks for our current study, we adopted an initial redshift selection of $0.45 < z < 1.0$ with a gap between $0.515 < z < 0.575$ because at these redshifts the Hβ and [OIII] lines lie in the region of the strong and highly structured atmospheric A-band absorption feature. The parent CFRS sample was originally defined to have $I_{AB} < 22.5$ to a faint isophotal limit (Lilly et al. 1995). As shown in the upper panel of Fig. 2, this corresponds over the redshift range of interest to selection of roughly $L^*$ galaxies, with a lower limit increasing from $M_{B,AB} < -20$ at $z \sim 0.5$ to $M_{B,AB} < -21$ at $z \sim 0.9$, where there is a more or less direct correspondence between the observed $I$-band and the rest-frame $B$-band.

Ideally, we would like an Hα or Hβ selected sample. However, these lines had not previously been observed for most of the redshift range of interest because of the spectral range of the original CFRS spectroscopy. Accordingly, in order to observe galaxies that might be expected to have sufficiently strong Hβ for useful study, we selected galaxies that had been observed in the original CFRS to have an [OII] λ3727 flux, $f_{3727} \geq 10^{-16}$ erg s$^{-1}$ cm$^{-2}$ for galaxies at $z < 0.92$. Empty spaces in the mask design were filled, if possible, with higher redshift objects $z > 0.92$ with $f_{3727} \geq 3 \times 10^{-16}$ erg s$^{-1}$ cm$^{-2}$ and at lower redshifts $z < 0.45$ with no flux limit.

The middle panel shows the luminosity of [OII] λ3727 from the new observations compared with the limits implied by this flux selection. A few galaxies now lie below the nominal line because of variations in the observational setup, e.g., choice of slit width, slit positioning etc., leading to lower [OII] λ3727 fluxes than in the original CFRS spectra. It should be noted that CFRS galaxies with $f_{3727} \geq 10^{-16}$ erg s$^{-1}$ cm$^{-2}$ produce more than 90% of the [OII] λ3727 luminosity density observed in the CFRS at $0.5 < z < 0.9$. The CFRS itself is estimated to produce about 50% of the total ultraviolet luminosity density of the Universe at these redshifts (Lilly et al. 1996). Thus, the present sample is presumably associated with of order one half of the overall [OII] λ3727 luminosity density in the Universe at these epochs and thus, excluding very heavily obscured sources, with the same fraction of the star-formation activity.

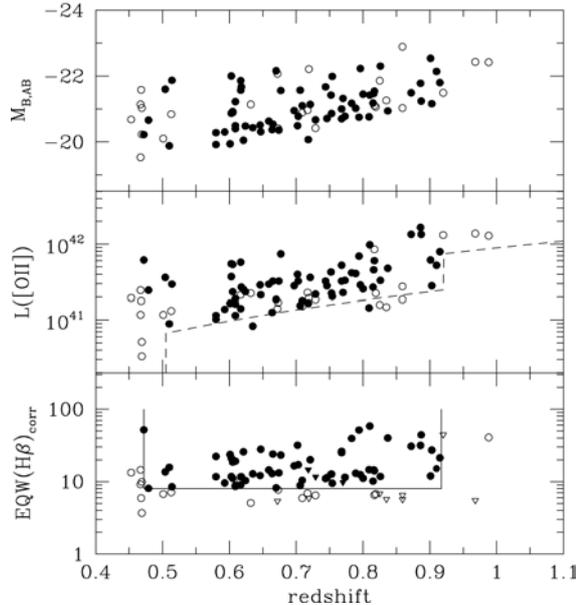

Fig. 2. The selection of the final statistically complete sample of 66 galaxies studied in this paper (solid symbols) from the sample of 94 galaxies observed (all symbols) as described in the text. Triangles represent limits. The upper panel illustrates the original CFRS selection of $I_{AB} < 22.5$, leading to a luminosity limit in the rest-frame B-band that gradually increases with redshift. The middle panel illustrates the selection in [OII] λ3727 flux density of $10^{-16}$ erg s$^{-1}$ cm$^{-2}$ over most of the redshift range, representing 90% of the [OII] luminosity in the CFRS at these epochs. The lowest panel shows the final cut in redshift space and rest-frame absorption-corrected Hβ equivalent width.

As discussed in Paper 1, there are many advantages to imposing a *post facto* selection in Hβ rather than [OII] λ3727. The Hβ strength is less affected by extinction, cooling and ionization effects than [OII] λ3727, and it is a more direct measure of the star-formation rate. Furthermore, since Hβ is generally the weakest of the lines it usually dominates the uncertainty in $R_{23}$.

In Paper 1, we applied a cut in Hβ luminosity. In the current work, we have applied a selection in the rest-frame equivalent width of Hβ. This has the operational advantage that the correction to Hβ for underlying stellar absorption features is a correction of Hβ equivalent width. The equivalent width is also a rough indicator of the star-formation rate per unit stellar mass, which may be a more interesting parameter in the context of the evolution of a galaxy than the straight star-formation rate.

As shown in the bottom panel of Fig. 2, we have therefore applied two additional cuts in addition to the original $I_{AB} < 22.5$ and $f_{3727} \geq 10^{-16}$ erg s$^{-1}$ cm$^{-2}$ selection criteria. First, the rest-frame equivalent width of H$\beta$, after the correction of +3 Å for stellar absorption, is required to be EQW$_0$(H$\beta$) > 8 Å. This limits the effect of the uncertainty in the absorption correction ($\pm 2$ Å) to less than 25% in $R_{23}$. It also has the added convenient effect of eliminating most of the objects for which only an upper limit to the H$\beta$ flux was obtained, although it should be noted that three objects with limits above this cut have been retained since they could be members of the sample if their fluxes were close to the limit. Second, we have applied precise redshift limits of $0.470 < z < 0.917$, the low limit corresponding to the appearance of [OII] $\lambda$3727 in the 5500 Å filter used in the later observations and the high limit to the entry of H$\beta$ into a particularly bad region of the OH forest. With these additional limits applied, 66 galaxies remain in the "statistically complete" sample.

A legitimate question is whether the final "statistically complete" H$\beta$-selected sample is still biased relative to its [OII] $\lambda$3727 properties. Fig 3 shows the H$\beta$ equivalent width and [OII] $\lambda$3727/H$\beta$ ratio for the galaxies with $0.470 < z < 0.917$. Lines of constant [OII] $\lambda$3727 strength cut diagonally across the diagram, as shown. The concern would be whether we are losing galaxies in the final sample with low [OII] $\lambda$3727/H$\beta$ ratio as the H$\beta$ strength drops (since they could potentially have been below the $f_{3727}$ selection threshold). There is no evidence from Fig 3 that this is the case: the galaxies that are excluded by our H$\beta$ cut in fact span the full range of [OII] $\lambda$3727/H$\beta$ ratio (although some have only lower limits to this ratio) and the lower envelope of the distribution of points in [OII] $\lambda$3727/H$\beta$ actually falls as the H$\beta$ strength drops towards our imposed threshold. We will therefore adopt this "statistically complete" sample for the remainder of the paper as unbiased with respect to the [OII] $\lambda$3727 properties. Line fluxes for [OII] $\lambda$3727, H$\beta$ and [OIII] $\lambda$5007 are listed for these 66 galaxies in Table 1, together with the rest-frame equivalent widths of the first two lines. It should be noted that the listed uncertainties in Table 1 include those terms that are relevant for the relative line fluxes and do not include the 15% uncertainties in absolute fluxes.

### 3.2 *Comparison samples*

As in Paper 1, our primary source for a local comparison sample is that of Jansen et al. (2000, hereafter JFFC), who observed the Nearby Field Galaxy Sample (NFGS) of about 200 galaxies selected from the first CfA redshift catalogue. This sample is claimed to include galaxies of all morphological types and to span 8 magnitudes in luminosity and a broad range of environments. Most crucially for the present work, the spectra are integrated over most of the luminous parts of the galaxies and should thus be similar to the unresolved spectra of high redshift galaxies obtained here. Following JFFC, a correction for stellar absorption of 1 Å has been applied to the NFGS H$\beta$ equivalent widths, and a corresponding increase to the H$\beta$ fluxes.

In principle, one could anticipate that precisely the same selection criteria could be applied to the NFGS sample as we have applied at high redshifts. However, Fig. 4, which shows the $M_{B,AB}$-EQW$_0$(H$\beta$) plane for our sample and that of the NFGS, illustrates a difficulty encountered in this and most other studies of the evolution of galaxies with redshift. It can be seen that there are very few local analogues of the galaxies with the high (~$L^*$) luminosities and high H$\beta$ equivalent widths that dominate the CFRS at high

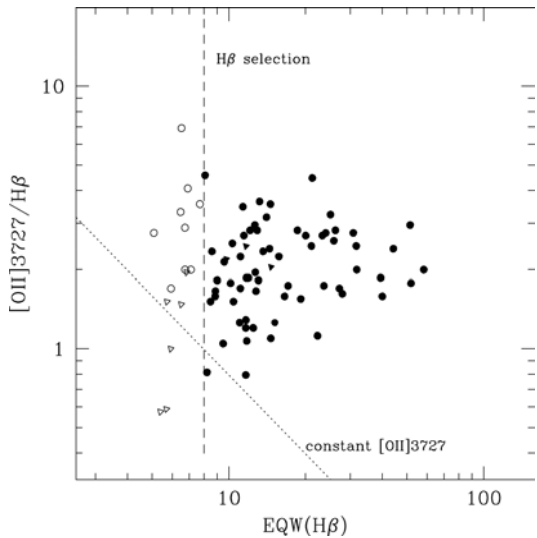

*Fig. 3. The emission line ratio [OII] $\lambda$3727/H$\beta$ is plotted against the rest-frame absorption-corrected H$\beta$ equivalent width for galaxies with $0.47 < z < 0.92$ to examine whether the final statistically complete sample (solid symbols) is biased with respect to the [OII] $\lambda$3727 properties on account of the original selection in [OII] $\lambda$3727 flux (middle panel of Fig. 2). The galaxies exhibit only a modest range in [OII] $\lambda$3727/H$\beta$ and there is no evidence of having lost galaxies with low [OII] $\lambda$3727/H$\beta$ at low H$\beta$ strengths and in fact the lower locus of points reflects a trend in the opposite direction. Three galaxies with limits to the H$\beta$ flux that are above the H$\beta$ selection threshold are retained in the sample, these are represented as solid triangles in this Figure. Other galaxies with limits below the threshold are shown as open triangles.*

TABLE 1

| CFRS | $z$ | [OII] 3727 $f^a$ | EQW$_0^b$ | Hβ $f^a$ | EQW$_0^b$ | [OIII] 5007 $f$ |
|---|---|---|---|---|---|---|
| 03.0062 | 0.826 | 7.30 ± 0.33 | 15 | 6.85 ± 1.34 | 12 | < 1.00 |
| 03.0085 | 0.609 | 8.80 ± 0.33 | 29 | 5.69 ± 0.63 | 19 | 7.70 ± 0.20 |
| 03.0125 | 0.789 | 10.10 ± 0.26 | 30 | 5.58 ± 0.92 | 13 | 1.30 ± 0.66 |
| 03.0145 | 0.603 | 17.70 ± 0.66 | 49 | 7.23 ± 0.76 | 21 | 8.80 ± 0.40 |
| 03.0261 | 0.697 | 9.50 ± 0.59 | 60 | 5.98 ± 0.75 | 17 | 14.40 ± 0.66 |
| 03.0327 | 0.609 | 5.30 ± 0.20 | 29 | 3.33 ± 0.82 | 9 | 1.90 ± 0.40 |
| 03.0488 | 0.605 | 25.50 ± 0.66 | 49 | 9.06 ± 1.03 | 19 | 13.80 ± 0.33 |
| 03.0570 | 0.646 | 11.60 ± 0.33 | 39 | 4.12 ± 0.73 | 12 | 6.90 ± 0.33 |
| 03.0595 | 0.605 | 11.10 ± 0.33 | 28 | 6.58 ± 1.21 | 11 | 3.70 ± 0.33 |
| 03.0599 | 0.479 | 20.50 ± 0.66 | 62 | 4.46 ± 1.14 | 8 | 9.00 ± 0.40 |
| 03.0879 | 0.601 | 8.00 ± 0.40 | 49 | 4.58 ± 0.47 | 24 | 2.40 ± 0.26 |
| 03.0999 | 0.706 | 5.00 ± 0.20 | 27 | 3.02 ± 0.71 | 9 | < 2.00 |
| 03.1016 | 0.702 | 13.10 ± 0.46 | 75 | 6.51 ± 0.49 | 32 | 7.50 ± 0.53 |
| 03.1112 | 0.768 | 8.50 ± 0.33 | 56 | 3.05 ± 0.46 | 26 | 7.00 ± 0.33 |
| 03.1138 | 0.768 | 13.80 ± 0.66 | 89 | 4.32 ± 0.48 | 25 | 6.80 ± 0.33 |
| 03.1309 | 0.617 | 25.70 ± 0.33 | 20 | 14.10 ± 3.20 | 9 | 5.90 ± 0.40 |
| 03.1349 | 0.617 | 6.40 ± 0.26 | 17 | 8.08 ± 1.44 | 12 | 7.50 ± 0.66 |
| 03.1367 | 0.703 | 10.50 ± 0.40 | 36 | 6.06 ± 0.74 | 17 | 9.20 ± 0.40 |
| 03.1375 | 0.635 | 3.50 ± 0.20 | 28 | 2.09 ± 0.38 | 13 | 2.50 ± 0.26 |
| 03.1534 | 0.794 | 17.00 ± 0.33 | 106 | 5.73 ± 0.35 | 5 | 10.80 ± 0.26 |
| 03.9003 | 0.618 | 12.00 ± 0.20 | 23 | 10.10 ± 1.78 | 12 | 2.60 ± 0.40 |
| 10.0478 | 0.752 | 11.80 ± 0.53 | 38 | 4.06 ± 0.72 | 13 | 7.60 ± 0.59 |
| 10.1116 | 0.709 | 6.00 ± 0.30 | 24 | 3.60 ± 0.70 | 9 | < 2.00 |
| 10.1213 | 0.815 | 6.19 ± 0.18 | 27 | 3.45 ± 1.61 | 10 | 3.00 ± 1.50 |
| 10.1608 | 0.729 | 7.21 ± 0.35 | 41 | < 2.64 | 11 | < 2.40 |
| 10.1925 | 0.783 | 10.56 ± 0.44 | 56 | 5.68 ± 0.56 | 39 | 5.92 ± 1.97 |
| 10.2183 | 0.910 | 9.20 ± 0.46 | 31 | 7.34 ± 0.66 | 15 | 6.90 ± 0.66 |
| 10.2284 | 0.773 | 8.53 ± 0.26 | 32 | 3.21 ± 1.53 | 11 | < 7.80 |
| 10.2418 | 0.796 | 7.05 ± 0.88 | 22 | 5.82 ± 0.98 | 12 | < 3.00 |
| 10.2428 | 0.872 | 26.01 ± 0.36 | 72 | 9.44 ± 0.93 | 31 | 28.34 ± 0.50 |
| 10.2519 | 0.718 | 4.05 ± 0.26 | 40 | < 2.70 | < 22 | < 2.40 |
| 10.2548 | 0.770 | 5.98 ± 0.44 | 59 | < 2.70 | < 10 | 2.54 ± 0.39 |
| 14.0072 | 0.621 | 11.20 ± 0.26 | 88 | 4.30 ± 0.42 | 26 | 10.80 ± 0.26 |
| 14.0129 | 0.903 | 5.00 ± 0.20 | 36 | 2.92 ± 0.34 | 27 | < 2.70 |
| 14.0217 | 0.721 | 11.09 ± 0.44 | 52 | 4.20 ± 0.60 | 20 | 5.80 ± 0.28 |
| 14.0272 | 0.670 | 6.90 ± 0.26 | 10 | 8.35 ± 2.05 | 8 | 1.60 ± 0.40 |
| 14.0393 | 0.603 | 26.06 ± 0.35 | 28 | 14.13 ± 2.44 | 12 | 10.19 ± 0.57 |
| 14.0497 | 0.800 | 6.20 ± 0.26 | 33 | 2.74 ± 0.63 | 11 | 2.90 ± 0.66 |
| 14.0538 | 0.810 | 22.69 ± 0.35 | 122 | 11.25 ± 0.40 | 58 | 17.80 ± 0.60 |
| 14.0605 | 0.837 | 10.12 ± 0.18 | 71 | 6.40 ± 0.76 | 40 | 12.55 ± 0.50 |
| 14.0725 | 0.580 | 5.30 ± 0.26 | 73 | 4.74 ± 0.63 | 22 | 2.10 ± 0.20 |
| 14.0779 | 0.580 | 6.00 ± 0.26 | 31 | 4.71 ± 1.04 | 12 | 2.20 ± 0.26 |
| 14.0818 | 0.901 | 10.87 ± 0.18 | 29 | 5.90 ± 1.07 | 12 | 4.88 ± 2.44 |
| 14.0848 | 0.664 | 4.70 ± 0.20 | 47 | 1.69 ± 0.37 | 13 | 2.10 ± 0.26 |
| 14.0972 | 0.677 | 26.52 ± 0.44 | 60 | 9.80 ± 0.86 | 23 | 16.60 ± 0.47 |
| 14.0985 | 0.809 | 3.35 ± 0.35 | 19 | 3.05 ± 0.51 | 15 | 3.80 ± 0.40 |
| 14.1087 | 0.659 | 11.34 ± 0.27 | 36 | 4.68 ± 0.65 | 14 | 8.31 ± 0.09 |
| 14.1126 | 0.746 | 7.90 ± 0.26 | 43 | 2.31 ± 0.45 | 11 | 5.30 ± 0.26 |
| 14.1189 | 0.753 | 6.15 ± 0.18 | 25 | 3.21 ± 0.69 | 13 | < 5.00 |
| 14.1190 | 0.754 | 5.60 ± 0.33 | 13 | 5.41 ± 1.19 | 9 | < 1.50 |
| 14.1258 | 0.647 | 8.51 ± 0.18 | 64 | 5.34 ± 0.39 | 28 | 7.38 ± 0.19 |
| 14.1386 | 0.744 | 9.32 ± 0.18 | 22 | 7.39 ± 1.42 | 11 | 2.79 ± 0.29 |
| 14.1466 | 0.674 | 11.80 ± 0.66 | 66 | 3.23 ± 0.53 | 13 | 5.20 ± 0.26 |
| 14.9705 | 0.609 | 7.62 ± 0.53 | 28 | 3.23 ± 0.80 | 9 | 2.47 ± 0.29 |
| 22.0274 | 0.504 | 27.00 ± 0.33 | 24 | 10.81 ± 1.71 | 14 | 36.20 ± 1.0 |
| 22.0322 | 0.915 | 13.70 ± 0.26 | 60 | 3.03 ± 0.72 | 21 | 7.70 ± 0.53 |
| 22.0417 | 0.593 | 6.90 ± 0.26 | 32 | 3.20 ± 0.70 | 10 | 2.60 ± 0.26 |
| 22.0429 | 0.624 | 10.20 ± 0.33 | 44 | 4.09 ± 0.86 | 10 | 3.60 ± 0.33 |
| 22.0576 | 0.887 | 25.00 ± 0.33 | 85 | 10.51 ± 0.54 | 44 | 18.90 ± 0.66 |
| 22.0599 | 0.886 | 31.00 ± 0.33 | 73 | 12.48 ± 0.86 | 32 | 11.70 ± 1.32 |
| 22.0770 | 0.816 | 10.50 ± 0.26 | 46 | 2.90 ± 0.52 | 15 | 4.00 ± 0.33 |
| 22.0919 | 0.472 | 51.00 ± 0.66 | 79 | 27.04 ± 1.33 | 52 | 143.00 ± 0.66 |
| 22.1119 | 0.514 | 20.80 ± 0.46 | 16 | 13.62 ± 3.24 | 8 | 15.00 ± 1.98 |
| 22.1313 | 0.817 | 13.60 ± 0.40 | 54 | 4.32 ± 0.90 | 14 | 8.20 ± 0.40 |
| 22.1350 | 0.510 | 6.40 ± 0.33 | 37 | 2.84 ± 0.41 | 16 | 2.30 ± 0.26 |
| 22.1528 | 0.665 | 12.80 ± 0.40 | 66 | 4.54 ± 0.40 | 24 | 8.50 ± 0.40 |

Notes:
(a) Line flux in units of $10^{-17}$ erg s$^{-1}$ cm$^{-2}$
(b) Rest-frame equivalent width in units of ≈.

redshifts. The uncertainty about the present day descendents of the star-forming galaxies that dominate the CFRS at $z > 0.5$ is of course a major motivation for the present study.

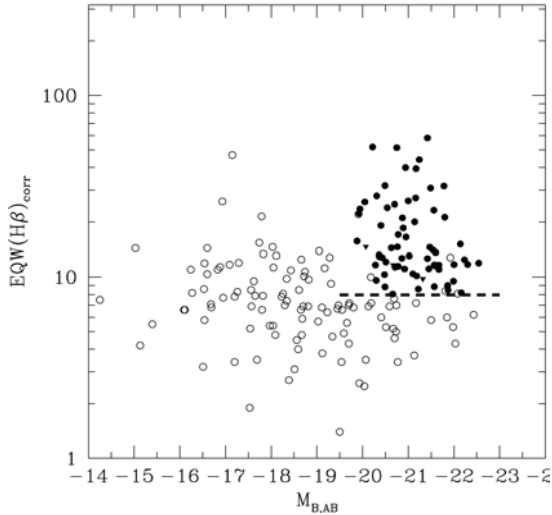

*Fig. 4. The rest-frame absorption-corrected Hβ equivalent width vs. absolute AB magnitude in the B-band for the 66-galaxy CFRS sample (solid symbols) and the NFGS sample of local galaxies (open symbols). The three triangles represent galaxies for which only an upper limit (above the selection limit) to the Hβ equivalent width was obtained. The diagram illustrates the well-known fact that the CFRS is dominated at high redshifts by galaxies that are much more luminous than local counterparts with the same properties. Thus the nature of these galaxies and their evolutionary descendents must be determined astrophysically, as is attempted in this paper.*

In addition to the NFGS sample, we also consider comparisons with the extensive extragalactic HII region data of van Zee et al. (1998) which span a wide range of metallicities and galactocentric radii, the relatively low redshift ultraviolet selected galaxy sample of Contini et al. (2002) and the handful of Lyman Break Galaxies (LBG) that have been observed at $z > 2.5$ (Pettini et al. 2001, Kobulnicky & Koo 2000).

*(a) The $R_{23}$-$O_{32}$ diagram*

Fig 5 shows the diagnostic diagram for the 66-galaxy CFRS sample constructed from $R_{23}$ and the oxygen ionization parameter $O_{32}$:

(2) $$O_{32} = \frac{f_{[OIII]4959,5007}}{f_{[OII]3727}}$$

Also plotted on this diagram is the McGaugh (1991) calibration of $R_{23}$ and $O_{32}$ in terms of [O/H] metallicity. McGaugh et al. used a solar value of [O/H] = log (O/H)+12 = 8.90 whereas Allende Prieto et al (2001) give a preferred solar value of [O/H] = 8.68. To avoid confusion should the preferred solar value continue to evolve, our discussion below is based wherever possible on [O/H] rather than metallicities relative to solar.

The lines of constant [O/H], which are only weakly dependent on $O_{32}$, first move rightwards across the diagram to higher $R_{23}$ as the metallicity increases from very low values. Then for [O/H] greater than about 8.4, the lines double back leftwards to smaller $R_{23}$, due to oxygen cooling effects. The precise turnaround point depends on the ionization parameter but is in the range 8.3-8.5 for a wide range of ionization parameters (see Fig 12 of McGaugh et al 1991). The effects of reddening on $R_{23}$ depend on $O_{32}$ (but not on $R_{23}$ itself) and are shown for $E_{B-V} = 0.3$ (i.e., $A_V \sim 1$) on Fig 5. It should be noted that increased extinction moves points away from the turnaround zone towards higher or lower metallicities. Correspondingly, over-correction for reddening will make objects crowd towards the turnaround area.

Fig 6 compares the CFRS data with other comparison samples. Panel (a) shows the McGaugh (1991) calibration from Fig 5. Panels (b)-(d) show the three low redshift comparison samples. The van Zee HII region sample was corrected (in the original paper) for reddening. Reddening corrections have not been applied to the remainder of the samples. The few $z \sim 3$ LBGs from Pettini et al. (2001) are represented in panel (e) while our own CFRS data are reproduced from Fig. 5 as panel (f). The solid line reproduced on all panels is a fit, by eye, to the ridgeline of the distribution of NFGS galaxies in panel (c).

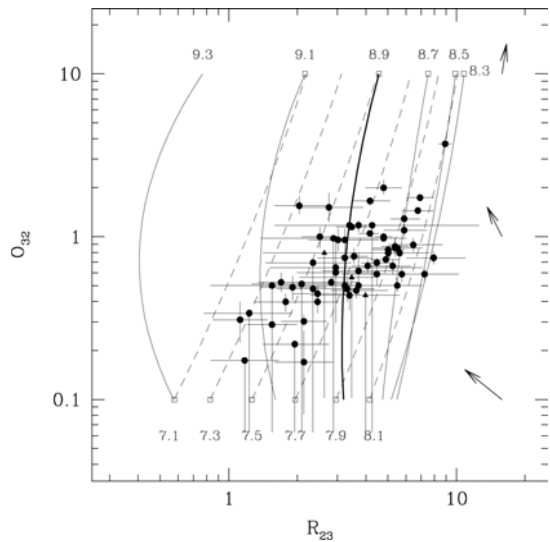

*Fig. 5. The $R_{23}$ vs. $O_{32}$ diagram for the complete sample considered in this paper. Also show is the McGaugh (1991) calibration showing how lines of constant [O/H] first move across the diagram rightwards as the metallicity increases (dashed lines) and then reverse at around [O/H] ~ 8.4 before returning leftwards as the metallicity increases further (solid lines) through the Solar value (heavy line). The lines are labelled with [O/H]. The effect of reddening for an $E_{B-V} = 0.3$ is shown for different values of $O_{32}$*

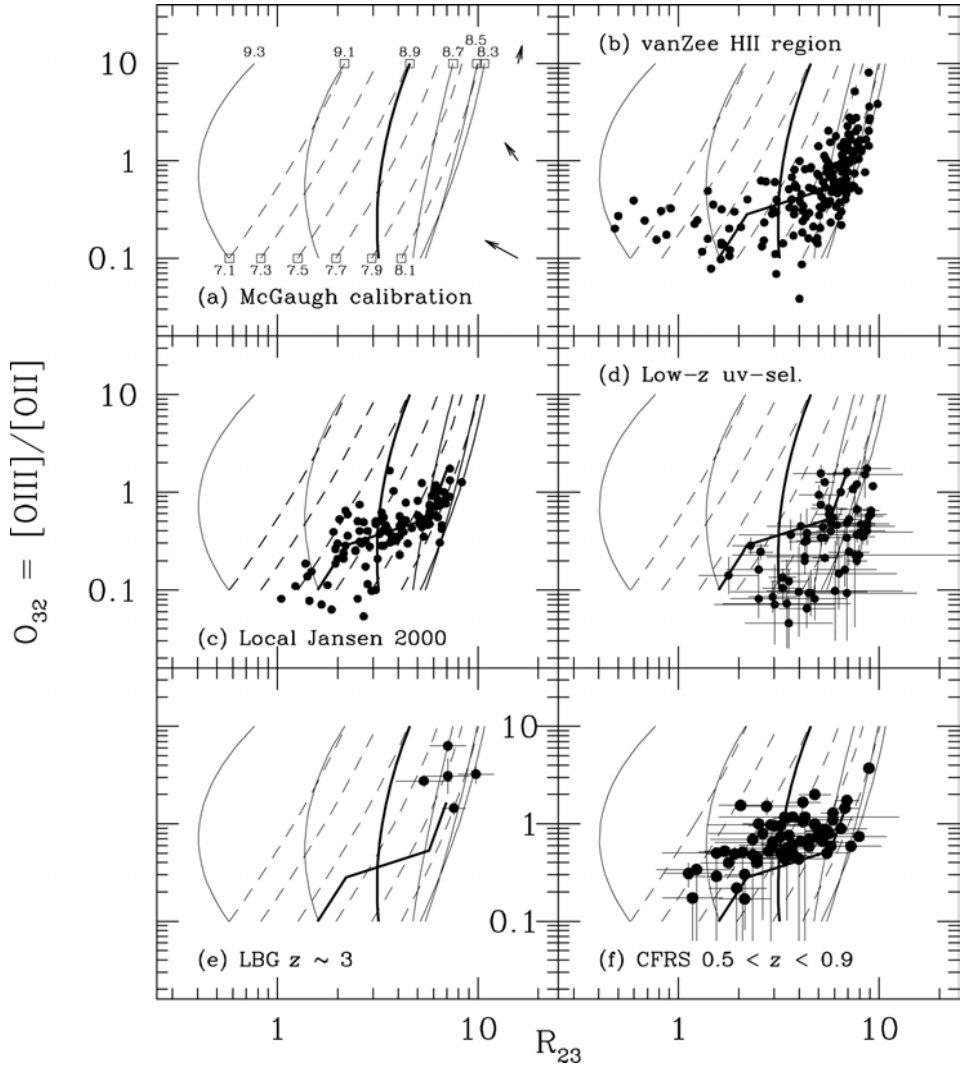

*Fig. 6    As for Fig. 5, but comparing the data of this paper with other relevant data sets. Panel (a) The McGaugh (1991) [O/H] calibration as in Fig 5. Panel (b) Data points for extragalactic HII regions from van Zee et al. (1998); Panel (c) data points for the NFGS B-selected local galaxy sample (see text); (d) data points for the low redshift sample of ultraviolet selected galaxies from Contini et al. (2002); Panel (e) data points from 5 Lyman break galaxies from Pettini et al. and (2001); Panel (f) data points from the present study of 66 CFRS galaxies with $0.47 < z < 0.92$, as in Fig. 5. Only the van Zee data in (b) have been de-reddenned - the remainder are as observed. The line in panels (b)-(f) is a fit by eye to the ridge line of the NFGS data in (c). Comparison of the panels reveals a closer similarity between the spectra of the CFRS galaxies and the local NFGS sample than with the other samples. The small offset between CFRS and NFGS is likely caused by a slightly higher level of extinction in the CFRS sample compared with the NFGS sample which spans a much broader range of luminosities extending down to $M_B \sim -14$ (see Fig. 9).*

The CFRS galaxies occupy a similar area on the $R_{23}$-$O_{32}$ plane to that of the NFGS sample. The small displacement up and to the left, noticed in Paper 1, persists in the larger data set. As noted in Paper 1, this may well be due to greater extinction in the CFRS (neither sample has been reddening corrected on this figure). The offset would correspond to roughly $\Delta E_{B-V} \sim 0.2$. This is quite plausible: the NFGS sample spans a very large range in luminosity $-14 < M_{B,AB} < -23$ and, since there is a strong correlation between extinction and luminosity (see Fig 7 below), the median extinction in the NFGS sample, $<E_{B-V}> \sim 0.15$, would be expected to be somewhat lower than would be found in the brighter $M_{B,AB} < -20$ CFRS galaxies. Other possibilities could be an incorrect correction for underlying stellar absorption features in the NFGS sample, in which most galaxies have quite low

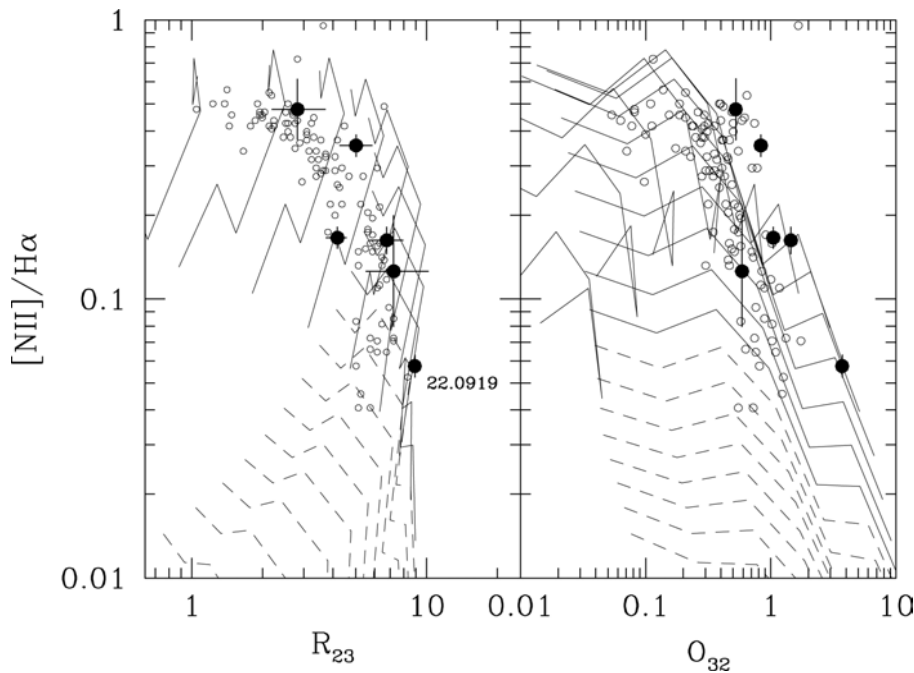

*Fig. 7. The [NII] λ6583/Hα ratio plotted against $R_{23}$ and $O_{32}$ for the 6 CFRS galaxies for which we have Hα and [NII] spectroscopy (solid points) and for the NFGS (open symbols). The lines indicate lines of constant [O/H] and varying ionization q from the models of Kewley and Dopita (2002). The lines are spaced in 0.1 dex in [O/H] and solid for [O/H] ≥ 8.4 are solid and dashed for [O/H] < 8.4 (as in Figs. 5 and 6a). The data show that all of the CFRS galaxies observed so far, and all but one of the NFGS, are consistent with [O/H] > 8.4, i.e. of being on the upper high metallicity branch of the $R_{23}$-[O/H] degeneracy. It should be noted that the CFRS galaxies targeted for Hα were preferentially selected to have high $R_{23}$*

equivalent widths of Hβ emission - see Fig. 4. Finally it could be a real physical effect.

Apart from this effect, Fig. 6 demonstrates that the CFRS spectra at $0.47 < z < 0.92$ generally exhibit line ratios that are broadly similar to those found in the best low redshift analogue sample, the NFGS sample, noting that that sample spans a very broad range of luminosity ñ14 $< M_{B,AB} <$ -23 whereas the CFRS sample is necessarily restricted to $M_{B,AB} <$ -20. The CFRS points are a little higher in the $O_{32}$ ratio at a given $R_{23}$ but the effect is small compared with the larger differences with the other comparison samples represented on Fig. 6.

In particular it is quite striking that only one of the 66 CFRS galaxies (22.0919 at $z = 0.472$, also observed by Hammer et al 2001) has $O_{32} > 2$, the maximum seen in the NFGS sample, and the vast majority have $O_{32} < 1$, implying $f_{3727} > f_{4959} + f_{5007}$. As will be discussed below, this is in stark contrast to the available data on the LBG galaxies at $z > 2.5$ (Pettini et al. 2001) where currently all have $O_{32} > 1$ and 4 of 5 have $O_{32} > 2$. The two $z = 1.9$ gravitationally-lensed galaxies observed by Lemoine-Busserole et al (2003) also both have $O_{32} > 2$.

The degeneracy between high and low metallicity solutions for $R_{23}$ can be broken using the [NII] λ6583 line. In addition to the J-band spectroscopy of four objects, our spectral range enabled us also to cover Hα and [NII] for two of the lowest redshift objects in the sample. The locations of these 6 objects on the [NII] λ6583/Hα vs. $R_{23}$ and [NII] λ6583/Hα vs. $O_{32}$ diagrams are shown in Fig 7. The [NII] λ6583/Hα ratio is quite strongly dependent on the ionization parameter (see e.g. Kewley and Dopita 2002), but has the advantage that it does not require a photometrical calibration between the infrared and optical data. Furthermore, it is insensitive to reddening. The dependence on the ionization parameter can be circumvented by simultaneously considering both $R_{23}$ and $O_{32}$ as on Fig 7. We show there the lines of constant metallicity and varying $q$ parameter that are derived from the models and polynomial fits of Kewley and Dopita (2002). All of the available objects are consistent with lying at [O/H] > 8.4, i.e. broadly speaking, above the turn-around point in the $R_{23}$-Z degeneracy.

We have also plotted on Fig 7 the NGFS sample. For all but two of these galaxies the $R_{23}$, $O_{32}$ and [NII]/Hα ratios indicate also that [O/H] > 8.4.

It should be noted that objects expected to lie in the turn-around region were preferentially targeted for the infrared program and so the distribution of CFRS galaxies on Fig 7 is not representative of the sample as a whole.

*(b) The $R_{23}$- $M_{B,AB}$ diagram*

In Paper 1, we presented the $R_{23}$-$M_B$ diagram and suggested that this too remained remarkably similar to that seen at low redshifts. The variation of $R_{23}$ with blue absolute magnitude $M_{B,AB}$ in the new combined data set is shown in Fig. 8. In order to make the most meaningful comparison, the effect of differing ionization levels $O_{32}$ has been removed by applying a small correction to $R_{23}$ so that the plotted value are those that would be obtained for $O_{32} = 1$ (i.e. log $O_{32} = 0$). To do this, the galaxies are slid up, or down, the curves of constant metallicity in the $R_{23}$-$O_{32}$ plane (Figs. 5 and 6) from the observed ($R_{23}$,$O_{32}$) value to the intersection with the log $O_{32} = 0$ line. The slope of the constant metallicity curves at a given observed ($R_{23}$,$O_{32}$) point are different for the high and low metallicity branches (see Figs. 5 and 6), so it was assumed that all the galaxies lie on the high metallicity branch of the $R_{23}$ degeneracy in order to compute this small correction.

Fig. 8 immediately shows two things. First, and as reported in Paper 1, the majority of the CFRS galaxies still occupy the same area on the $M_{B,AB}$ and $R_{23}$ diagram as the low redshift NFGS sample with the same $M_{B,AB}$. On the other hand, with the increase in the sample size, we now find a small but significant number have the higher $R_{23} \geq 5$ values (i.e. lower metallicities) that are found only at significantly lower luminosities in the NFGS sample. The high $R_{23}$ values are not found in the most luminous NFGS objects in $M_B$.

In Paper 1, we identified two of fifteen objects in this category and we tentatively identified these as possible active galactic nuclei (AGN) on the basis of strong [NeIII] emission and thus concluded that there was no strong evidence for a departure from the low redshift $M_{B,AB}$-$R_{23}$ relation. However, our new *J*-band spectroscopy clearly establishes that these two galaxies are not AGN and the N[II]/Hα ratio (Fig. 7) indicates that they are in fact genuinely lower metallicity systems (see also Carollo et al. 2002 for discussion). Our present, larger, sample shows that objects of this type represent a significant minority of the population.

It should be noted that the three galaxies with lower limits to $R_{23}$ (i.e., upper limits to $f_{HB}$) have been retained in the diagram (as triangles). In fact, they would drop out of the sample if their $f_{HB}$ were much weaker, and this enables us to constrain $R_{23}$. Thus we can say that all of them would still be in the ì normalî part of the diagram if they are in the sample at all. We include them in the following discussion, though their exclusion would make little difference in our conclusions.

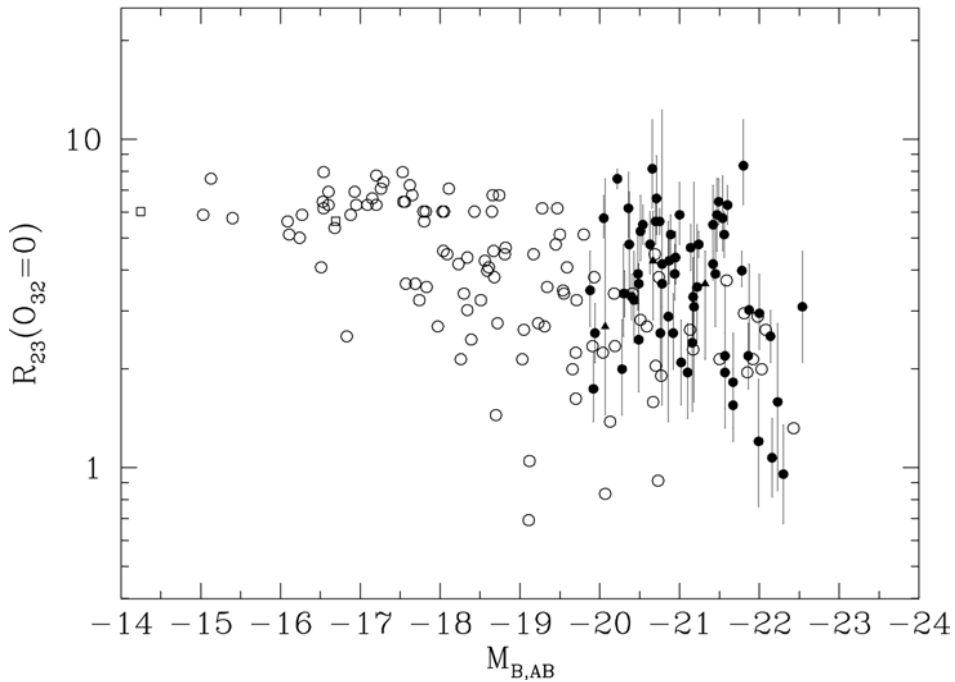

*Fig. 8. The values of $R_{23}$ (corrected to the value they would take for $O_{32}$=0 as described in the text) plotted against B-band absolute magnitude $M_{B,AB}$ for the CFRS 0.47 < z < 0.92 galaxies (solid symbols) and NFGS local sample (open symbols). Neither sample is corrected for reddening. The bulk of the CFRS galaxies have $R_{23}$ values that overlie those of galaxies of comparable luminosity in the local Universe. However, approximately 25% have higher values of $R_{23}$, i.e., suggestive of lower metallicities, which are found in the local sample only at lower luminosities.*

### 3.4 Metallicity estimations

We have calculated an [O/H] metallicity to each of the CFRS galaxies based on its $R_{23}$ value. We have simply applied the [O/H] calibration indicated on Figs 5, 6 to all of our galaxies, to the local NFGS sample, and to the available LBG galaxies. In particular, we ignore effects arising from variations in electron temperature or ionization parameter within galaxies, which, for compact galaxies, can alone lead to an underestimate of the metallicity by 0.1 dex (Kobulnicky et al. 1999). We also ignore the effects of radial metallicity gradients in larger galaxies assuming that the spectroscopy includes essentially all of the light from the galaxies in question (see the discussion in Kobulnicky et al. 1999). Undoubtedly, there are many effects that will introduce errors into our [O/H] measurements. Our philosophy is to focus on *relative* effects between similarly selected samples in the expectation that these are likely to be much more robust than attempts at absolute metallicity determinations.

Ideally, the spectra would be individually dereddened using $H\alpha/H\beta$ and the degeneracy broken by observations of [NII] $\lambda 6583$. Unfortunately, we have as yet infrared spectroscopy for only a few objects. We have not attempted to use the $H\alpha$ measurements for 11 of our galaxies that have recently been published by Tresse et al. (2002) since they were taken with a much larger slit and have a poorly defined continuum (which otherwise could have been used for a relative flux calibration). Likewise, we have not attempted to derive extinctions in the CFRS sample from $H\beta/H\gamma$ because of our poor signal to noise in $H\gamma$ and because of the difficulties of removing the underlying stellar absorption for the higher Balmer lines.

Fig. 9 shows the extinction derived from the Balmer $H\alpha/H\beta$ ratios in the low redshift NFGS sample as a function of *B*-band luminosity. These have been used to individually correct the $R_{23}$ and $O_{32}$ values of those galaxies in deriving the metallicities. The median value for the whole NFGS sample is $<E_{B-V}> = 0.15$, while that for the galaxies with $M_{B,AB} < -20$ is higher $<E_{B-V}> = 0.33$. In the absence of reliable individual extinction estimates for the CFRS galaxies, every CFRS spectrum has been dereddened assuming a uniform $E_{B-V} = 0.3 \pm 0.15$, propagating this extra uncertainty through to the uncertainty in [O/H]. This is the average for the NFGS sample of similar luminosities (comparable with the average

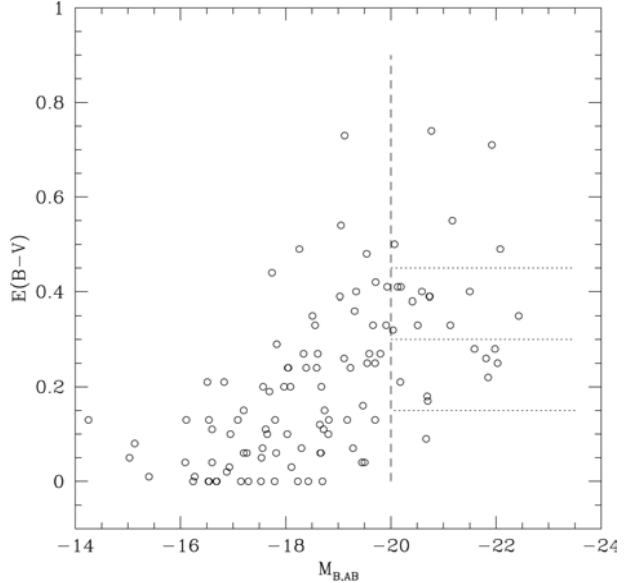

*Fig. 9. Extinction $E_{B-V}$ as derived from the $H\alpha/H\beta$ ratio plotted against B-band absolute magnitude for the NFGS local sample. The median extinction for the whole sample is $<E_{B-V}> = 0.15$ mag, only a half of that for the galaxies with $M_{B,AB} < -20$ ($<E_{B-V}> = 0.33$) which are likely the best analogues for the CFRS sample. This figure may explain the offset in the $R_{23}$-$O_{32}$ diagram between the CFRS and NFGS samples (Fig. 6c and f). The horizontal dotted lines show the reddening of $E_{B-V} = 0.3 \pm 0.15$ applied to the CFRS sample (see text).*

reddening of lower redshift CFRS galaxies in Tresse et al. 2002).

Table 2 lists the values of $R_{23}$ and $O_{32}$ derived from the line measurements and the $O_{32}$-corrected $R_{23,0}$. Subsequent columns list the limits to [O/H] for both the high and low branch solutions, as well as line and continuum luminosities and other properties of the galaxies as discussed below.

Fig. 10a shows the [O/H] estimates as a function of $M_{B,AB}$ for the CFRS and for NFGS assuming that all of the CFRS and all but two of the NFGS lie on the high metallicity branch of the $R_{23}$-[O/H] degeneracy (see Section 3.3(a) above). Fig. 10b shows the equivalent to Fig. 10a that would be obtained with the assumption that all the galaxies lie on the low metallicity branch. Fig 10c is the equivalent to Fig. 10a but has been computed for the limiting case of zero reddening in the CFRS sample.

The CFRS galaxies whose error bars in Fig. 10 extend to the turn-around point in $R_{23}$ clearly have poorly determined metallicities since their [O/H] limits would likely extend into the low-branch regime. Their error bars have been arbitrarily truncated at precisely [O/H] = 8.3 in these three figures (and subsequently in the Paper) to visually emphasize this fact. Likewise galaxies which had a nominal $R_{23}$ after dereddening that was greater than that of the turn-around point were assigned [O/H] =

8.3 with no error-bar. We stress that galaxies in the ambiguous reversal zone between 8.6 > [O/H] > 8.05 (i.e., between the dashed lines in Figs. 10a-c) have unconventional error bars and these figures should be interpreted accordingly.

Objects in the ambiguous zone could well have metallicities anywhere throughout the 8.05 < [O/H] < 8.6 range, particularly because the effect of reddening on [O/H] in this regime is disproportionately large. It should be noted that application of too large a reddening correction would drive objects towards the turn-around point itself.

Based on the available [N II] λ6583 spectroscopy (see Fig. 7) and comparison of Fig. 10a and 10b, we will assume that all of the CFRS galaxies are on the high-[O/H] branch of the $R_{23}$ degeneracy. Indeed, if this is not correct for galaxies well away from the turn-around point, then it would mean that we have found a substantial population of luminous galaxies with very low metallicities [O/H] < 8 at these moderate redshifts, which would be an interesting discovery. While we cannot rule this out until we have secured further *J*-band spectroscopy, we believe it is unlikely and adopt in this paper the conservative assumption that all CFRS galaxies are on the upper branch with [O/H] ≥ 8.4.

Fig. 10a strongly suggests that the majority (51/66 ~ 77%) of galaxies in our CFRS-based sample at 0.47 < *z* < 0.92 will not evolve into low metallicity ìdwarfî galaxies since they already have metallicities [O/H] > 8.6 and in most cases already have the [O/H] ~ 8.9 metallicities seen in luminous galaxies today. The fraction of lower metallicity galaxies with [O/H] < 8.6 depends quite strongly on the assumed reddening because the metallicity changes rapidly for small changes in $R_{23}$ near the turn-around region. As an example, if we apply no reddening correction at all to the CFRS sample (Fig. 10c), then the number of galaxies with nomimal [O/H] < 8.6 drops to only 3/66, or 5%. Since we would expect lower metallicity galaxies to exhibit less reddening across the sample, we suspect that reality lies somewhere between these extremes.

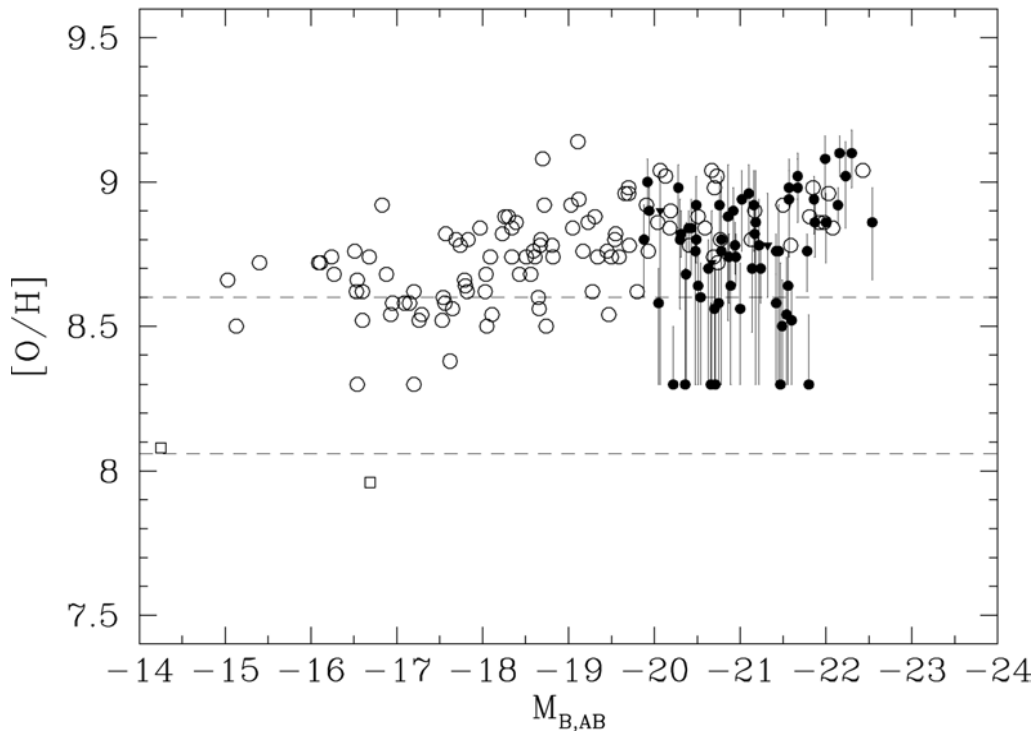

*Fig. 10 (a) [O/H] as a function of $M_{B,AB}$. [O/H] is estimated from $R_{23}$. Solid symbols are the 66 0.47 < z < 0.92 CFRS galaxies (triangles are limits) while open circles are NFGS galaxies with [O/H] > 8.4, open squares NFGS galaxies with [NeII]/Hα < 8.4, based on [NII] (see Fig. 7). It is assumed that all CFRS galaxies are on the high metallicity branch of the $R_{23}$-[O/H] degeneracy (see text and Fig 10(b)). The NFGS galaxies were individually dereddened based on their Hα/Hβ ratio while the CFRS were uniformly dereddened with $E_{B-V}$ = 0.3 (see Fig. 9). The metallicities of CFRS galaxies in the ì turn-aroundî region between the dashed lines are highly uncertain and the treatment of their location and error-bars is discussed in the text. 14 CFRS galaxies appear to have [O/H] < 8.6 and to thus exhibit metallicities that are well below those of NFGS galaxies of the same luminosities and that are found in the NFGS only at much lower luminosities.*

TABLE 2

| CFRS | z | Observed line ratios | | log $R_{23,0}$ | Upper branch [O/H][a] | | | Lower branch [O/H][a] | | | Line luminosity | | Photometry | | | Size |
|---|---|---|---|---|---|---|---|---|---|---|---|---|---|---|---|---|
| | | log $R_{23}$ | log $O_{32}$ | | max | best | min | min | best | max | H$\beta$[b] | [OII]3727[b] | $I_{AB}$ | $M_{B,AB}$ | $M_{J,AB}$ | log $r_{0.5}$[c] |
| 03.0062 | 0.826 | 0.07 ± 0.15 | < −0.76 | −0.02 | 9.18 | 9.10 | 8.98 | 7.36 | 7.56 | 7.78 | 41.49 | 41.52 | 20.96 | −22.30 | −23.61 | .... |
| 03.0085 | 0.609 | 0.53 ± 0.07 | 0.07 ± 0.03 | 0.52 | 8.90 | 8.84 | 8.76 | 7.64 | 7.74 | 7.86 | 41.09 | 41.28 | 22.00 | −20.40 | −21.64 | .... |
| 03.0125 | 0.789 | 0.33 ± 0.13 | −0.77 ± 0.20 | 0.32 | 9.04 | 8.94 | 8.76 | 7.66 | 7.84 | 8.08 | 41.35 | 41.61 | 22.08 | −21.02 | −22.27 | 0.61 |
| 03.0145 | 0.603 | 0.61 ± 0.07 | −0.18 ± 0.04 | 0.63 | 8.82 | 8.74 | 8.58 | 7.82 | 7.96 | 8.16 | 41.18 | 41.57 | 21.57 | −20.87 | −21.50 | 0.74 |
| 03.0261 | 0.697 | 0.68 ± 0.08 | 0.31 ± 0.06 | 0.64 | 8.82 | 8.74 | 8.64 | 7.72 | 7.84 | 7.98 | 41.25 | 41.45 | 21.75 | −20.95 | −22.87 | .... |
| 03.0327 | 0.609 | 0.37 ± 0.18 | −0.32 ± 0.11 | 0.39 | 9.02 | 8.92 | 8.74 | 7.50 | 7.72 | 8.00 | 40.86 | 41.06 | 21.89 | −20.49 | −21.92 | 0.4 |
| 03.0488 | 0.605 | 0.69 ± 0.06 | −0.14 ± 0.02 | 0.71 | 8.76 | 8.64 | (8.3) | 7.90 | 8.06 | (8.3) | 41.28 | 41.73 | 21.58 | −20.89 | −21.48 | 0.32 |
| 03.0570 | 0.646 | 0.70 ± 0.10 | −0.10 ± 0.04 | 0.72 | 8.76 | 8.64 | (8.3) | 7.88 | 8.06 | (8.3) | 41.01 | 41.46 | 22.07 | −20.51 | −21.00 | .... |
| 03.0595 | 0.605 | 0.39 ± 0.11 | −0.35 ± 0.06 | 0.41 | 8.98 | 8.90 | 8.76 | 7.58 | 7.74 | 7.96 | 41.14 | 41.37 | 21.46 | −20.92 | −22.08 | 0.15 |
| 03.0599 | 0.479 | 0.86 ± 0.12 | −0.23 ± 0.04 | 0.91 | 8.58 | (8.3) | (8.3) | 8.14 | (8.3) | (8.3) | 40.73 | 41.39 | 21.19 | −20.66 | −21.00 | 0.68 |
| 03.0879 | 0.601 | 0.39 ± 0.09 | −0.40 ± 0.07 | 0.41 | 8.98 | 8.90 | 8.80 | 7.62 | 7.78 | 7.94 | 40.98 | 41.22 | 22.48 | −19.94 | ..... | .... |
| 03.0999 | 0.706 | 0.32 ± 0.22 | < −0.29 | 0.34 | 9.08 | 8.94 | 8.74 | 7.38 | 7.64 | 7.98 | 40.97 | 41.19 | 21.22 | −21.57 | −23.69 | 0.81 |
| 03.1016 | 0.702 | 0.55 ± 0.06 | −0.12 ± 0.05 | 0.56 | 8.86 | 8.80 | 8.72 | 7.74 | 7.84 | 7.98 | 41.30 | 41.60 | 22.35 | −20.49 | −22.17 | 0.56 |
| 03.1112 | 0.768 | 0.77 ± 0.10 | 0.04 ± 0.04 | 0.77 | 8.72 | 8.56 | (8.3) | 7.92 | 8.12 | (8.3) | 41.06 | 41.51 | 22.03 | −21.00 | ..... | .... |
| 03.1138 | 0.768 | 0.72 ± 0.08 | −0.18 ± 0.04 | 0.75 | 8.72 | 8.56 | (8.3) | 7.96 | 8.16 | (8.3) | 41.21 | 41.72 | 22.33 | −20.70 | −21.51 | .... |
| 03.1309 | 0.617 | 0.33 ± 0.10 | −0.52 ± 0.04 | 0.34 | 9.02 | 8.94 | 8.82 | 8.60 | 7.74 | 7.92 | 41.50 | 41.76 | 20.62 | −21.86 | −23.79 | 0.95 |
| 03.1349 | 0.617 | 0.31 ± 0.12 | 0.19 ± 0.06 | 0.29 | 9.04 | 8.98 | 8.92 | 7.28 | 7.44 | 7.58 | 41.25 | 41.15 | 20.87 | −21.57 | −23.33 | 0.49 |
| 03.1367 | 0.703 | 0.57 ± 0.09 | 0.07 ± 0.04 | 0.56 | 8.88 | 8.80 | 8.72 | 7.68 | 7.80 | 7.96 | 41.27 | 41.51 | 22.08 | −20.78 | ..... | .... |
| 03.1375 | 0.635 | 0.51 ± 0.14 | −0.02 ± 0.07 | 0.51 | 8.94 | 8.84 | 8.68 | 7.58 | 7.76 | 8.00 | 40.70 | 40.92 | 22.15 | −20.43 | −22.18 | .... |
| 03.1534 | 0.794 | 0.74 ± 0.04 | −0.07 ± 0.02 | 0.75 | 8.70 | 8.58 | (8.3) | 8.00 | 8.12 | (8.3) | 41.37 | 41.84 | 22.45 | −20.75 | ..... | .... |
| 03.9003 | 0.618 | 0.19 ± 0.11 | −0.54 ± 0.08 | 0.19 | 9.10 | 9.02 | 8.94 | 7.44 | 7.60 | 7.78 | 41.35 | 41.43 | 20.77 | −21.67 | ..... | .... |
| 10.0478 | 0.752 | 0.73 ± 0.11 | −0.07 ± 0.06 | 0.74 | 8.76 | 8.58 | (8.3) | 7.90 | 8.10 | (8.3) | 41.16 | 41.63 | 21.57 | −21.42 | ..... | .... |
| 10.1116 | 0.709 | 0.29 ± 0.14 | < −0.66 | 0.29 | 9.06 | 8.96 | 8.78 | 7.56 | 7.76 | 8.02 | 41.07 | 41.25 | 21.71 | −21.10 | ..... | .... |
| 10.1213 | 0.815 | 0.47 ± 0.30 | −0.19 ± 0.20 | 0.49 | 9.04 | 8.86 | (8.3) | 7.42 | 7.78 | (8.3) | 41.18 | 41.43 | 22.07 | −21.18 | −22.89 | .... |
| 10.1608[d] | 0.729 | 0.42 − 0.75 | < −0.36 | 0.63 | 8.90 | 8.72 | (8.3) | 7.76 | 8.04 | (8.3) | 40.95 | 41.34 | 22.21 | −20.67 | −22.41 | .... |
| 10.1925 | 0.783 | 0.51 ± 0.12 | −0.13 ± 0.17 | 0.52 | 8.92 | 8.82 | 8.72 | 7.64 | 7.80 | 8.00 | 41.35 | 41.62 | 21.94 | −21.17 | −22.28 | .... |
| 10.2183 | 0.910 | 0.40 ± 0.08 | 0.00 ± 0.07 | 0.40 | 8.98 | 8.92 | 8.86 | 7.50 | 7.62 | 7.74 | 41.62 | 41.72 | 21.45 | −22.14 | −23.40 | .... |
| 10.2284 | 0.773 | 0.63 ± 0.45 | < 0.07 | 0.62 | 9.02 | 8.76 | (8.3) | 7.38 | 7.86 | (8.3) | 41.49 | 41.52 | 22.25 | −20.78 | −21.63 | .... |
| 10.2418 | 0.796 | 0.19 ± 0.24 | < −0.30 | 0.20 | 9.14 | 9.02 | 8.84 | 7.16 | 7.50 | 7.82 | 41.38 | 41.46 | 20.81 | −22.23 | −23.53 | .... |
| 10.2428 | 0.872 | 0.83 ± 0.07 | 0.16 ± 0.02 | 0.81 | 8.66 | 8.50 | (8.3) | 8.00 | 8.22 | (8.3) | 41.69 | 42.13 | 21.97 | −21.49 | −21.84 | .... |
| 10.2519[d] | 0.718 | 0.15 − 0.87 | < −0.10 | 0.43 | 9.04 | 8.90 | (8.3) | 7.38 | 7.68 | (8.3) | 40.91 | 41.22 | 22.49 | −20.07 | −20.67 | .... |
| 10.2548[d] | 0.770 | 0.31 − 0.64 | < −0.25 | 0.56 | 8.96 | 8.78 | 8.60 | 7.60 | 7.90 | 8.14 | 41.02 | 41.36 | 21.56 | −21.32 | −23.05 | .... |
| 14.0072 | 0.621 | 0.77 ± 0.07 | 0.11 ± 0.02 | 0.76 | 8.70 | 8.58 | (8.3) | 7.94 | 8.08 | (8.3) | 40.99 | 41.40 | 22.46 | −20.05 | −21.05 | .... |
| 14.0129 | 0.903 | 0.37 ± 0.19 | < −0.16 | 0.38 | 9.04 | 8.92 | 8.76 | 7.40 | 7.64 | 7.90 | 41.22 | 41.45 | 22.41 | −21.16 | −23.07 | .... |
| 14.0217 | 0.721 | 0.65 ± 0.07 | −0.16 ± 0.03 | 0.67 | 8.78 | 8.70 | 8.48 | 7.86 | 8.02 | 8.26 | 41.13 | 41.56 | 21.73 | −21.14 | ..... | .... |
| 14.0272 | 0.670 | 0.05 ± 0.12 | −0.51 ± 0.13 | 0.03 | 9.16 | 9.1 | 9.02 | 7.26 | 7.44 | 7.60 | 41.36 | 41.27 | 20.51 | −22.16 | −23.72 | .... |
| 14.0393 | 0.603 | 0.45 ± 0.12 | −0.28 ± 0.03 | 0.47 | 8.96 | 8.86 | 8.72 | 7.62 | 7.80 | 8.00 | 41.47 | 41.74 | 20.44 | −22.00 | −22.90 | 0.75 |
| 14.0497 | 0.800 | 0.57 ± 0.17 | −0.21 ± 0.10 | 0.59 | 8.92 | 8.76 | (8.3) | 7.68 | 7.92 | (8.3) | 41.06 | 41.41 | 21.69 | −21.45 | −23.14 | .... |
| 14.0538 | 0.810 | 0.62 ± 0.04 | 0.02 ± 0.02 | 0.62 | 8.82 | 8.76 | 8.68 | 7.78 | 7.88 | 7.98 | 41.69 | 41.99 | 21.82 | −21.42 | ..... | .... |
| 14.0605 | 0.837 | 0.62 ± 0.09 | 0.22 ± 0.03 | 0.59 | 8.86 | 8.78 | 8.68 | 7.68 | 7.80 | 7.96 | 41.48 | 41.68 | 22.44 | −20.94 | ..... | 0.58 |
| 14.0725 | 0.580 | 0.23 ± 0.11 | −0.28 ± 0.06 | 0.24 | 9.08 | 9.00 | 8.92 | 7.40 | 7.54 | 7.70 | 40.96 | 41.01 | 22.32 | −19.92 | −21.76 | 0.52 |
| 14.0779 | 0.580 | 0.28 ± 0.14 | −0.31 ± 0.08 | 0.30 | 9.06 | 8.98 | 8.84 | 7.42 | 7.60 | 7.82 | 40.95 | 41.06 | 22.01 | −20.28 | ..... | .... |
| 14.0818 | 0.901 | 0.47 ± 0.15 | −0.22 ± 0.20 | 0.49 | 8.98 | 8.86 | 8.66 | 7.56 | 7.80 | 8.06 | 41.52 | 41.79 | 21.02 | −22.54 | ..... | .... |
| 14.0848 | 0.664 | 0.65 ± 0.16 | −0.23 ± 0.07 | 0.68 | 8.86 | 8.68 | (8.3) | 7.80 | 8.04 | (8.3) | 40.65 | 41.10 | 22.30 | −20.37 | ..... | 0.32 |
| 14.0972 | 0.677 | 0.70 ± 0.07 | −0.08 ± 0.02 | 0.71 | 8.76 | 8.64 | (8.3) | 7.90 | 8.06 | (8.3) | 41.44 | 41.87 | 21.17 | −21.56 | −22.59 | 0.5 |
| 14.0985 | 0.809 | 0.44 ± 0.15 | 0.18 ± 0.09 | 0.41 | 9.00 | 8.92 | 8.80 | 7.40 | 7.58 | 7.78 | 41.12 | 41.16 | 22.45 | −20.76 | −22.09 | 0.44 |
| 14.1087 | 0.659 | 0.68 ± 0.13 | −0.01 ± 0.02 | 0.68 | 8.80 | 8.70 | (8.3) | 7.82 | 7.98 | (8.3) | 41.09 | 41.47 | 22.06 | −20.63 | −21.60 | 0.31 |
| 14.1126 | 0.746 | 0.81 ± 0.13 | −0.05 ± 0.04 | 0.82 | 8.70 | (8.3) | (8.3) | 7.98 | (8.3) | (8.3) | 40.91 | 41.45 | 22.26 | −20.71 | ..... | 0.72 |
| 14.1189 | 0.753 | 0.46 ± 0.29 | < −0.01 | 0.46 | 9.06 | 8.88 | 8.52 | 7.34 | 7.68 | 8.22 | 41.06 | 41.35 | 22.12 | −20.86 | −22.04 | 0.48 |
| 14.1190 | 0.754 | 0.09 ± 0.20 | < −0.47 | 0.08 | 9.16 | 9.08 | 8.94 | 8.20 | 7.46 | 7.72 | 41.29 | 41.31 | 20.99 | −21.99 | ..... | 0.63 |
| 14.1258 | 0.647 | 0.54 ± 0.07 | 0.06 ± 0.02 | 0.53 | 8.90 | 8.82 | 8.74 | 7.64 | 7.76 | 7.88 | 41.12 | 41.33 | 22.30 | −20.31 | −21.34 | 0.26 |
| 14.1386 | 0.744 | 0.25 ± 0.15 | 0.40 ± 0.06 | 0.26 | 9.08 | 8.98 | 8.86 | 7.42 | 7.62 | 7.82 | 41.41 | 41.41 | 21.28 | −21.67 | −23.15 | .... |
| 14.1466 | 0.674 | 0.76 ± 0.11 | −0.23 ± 0.04 | 0.79 | 8.70 | (8.3) | (8.3) | 8.02 | (8.3) | (8.3) | 40.95 | 41.51 | 22.34 | −20.36 | −20.94 | 0.45 |
| 14.9705 | 0.609 | 0.53 ± 0.18 | −0.36 ± 0.07 | 0.55 | 8.94 | 8.78 | (8.3) | 7.68 | 7.92 | (8.3) | 40.84 | 41.21 | 21.27 | −21.22 | ..... | .... |
| 22.0274 | 0.504 | 0.84 ± 0.06 | 0.24 ± 0.02 | 0.80 | 8.64 | 8.52 | (8.3) | 8.00 | 8.18 | (8.3) | 41.19 | 41.56 | 20.46 | −21.60 | −22.59 | .... |
| 22.0322 | 0.915 | 0.90 ± 0.13 | −0.13 ± 0.04 | 0.92 | 8.54 | (8.3) | (8.3) | 8.20 | (8.3) | (8.3) | 41.25 | 41.90 | 21.81 | −21.80 | −22.96 | .... |
| 22.0417 | 0.593 | 0.51 ± 0.15 | −0.30 ± 0.06 | 0.53 | 8.94 | 8.80 | 8.56 | 7.68 | 7.88 | 8.20 | 40.81 | 41.14 | 22.07 | −20.30 | −21.38 | .... |
| 22.0429 | 0.624 | 0.56 ± 0.12 | −0.33 ± 0.05 | 0.59 | 8.90 | 8.76 | (8.3) | 7.76 | 7.96 | (8.3) | 40.97 | 41.37 | 21.95 | −20.48 | −22.01 | .... |
| 22.0576 | 0.887 | 0.68 ± 0.04 | 0.00 ± 0.02 | 0.68 | 8.76 | 8.70 | 8.58 | 7.88 | 7.98 | 8.10 | 41.75 | 42.13 | 22.29 | −21.24 | ..... | 0.04 |
| 22.0599 | 0.886 | 0.57 ± 0.06 | −0.30 ± 0.05 | 0.60 | 8.82 | 8.76 | 8.62 | 7.84 | 7.96 | 8.12 | 41.83 | 42.22 | 21.74 | −21.78 | ..... | 0.35 |
| 22.0770 | 0.816 | 0.74 ± 0.11 | −0.30 ± 0.05 | 0.77 | 8.72 | (8.3) | (8.3) | 7.98 | (8.3) | (8.3) | 41.11 | 41.66 | 21.78 | −21.47 | −22.07 | .... |
| 22.0919 | 0.472 | 0.95 ± 0.03 | 0.57 ± 0.05 | 0.88 | 8.50 | (8.3) | (8.3) | 8.14 | (8.3) | (8.3) | 41.54 | 41.79 | 21.77 | −20.22 | ..... | .... |
| 22.1119 | 0.514 | 0.48 ± 0.14 | −0.02 ± 0.07 | 0.48 | 8.96 | 8.86 | 8.74 | 7.54 | 7.72 | 7.94 | 41.29 | 41.47 | 20.07 | −21.87 | ..... | .... |
| 22.1313 | 0.817 | 0.75 ± 0.13 | −0.10 ± 0.04 | 0.76 | 8.74 | 8.54 | (8.3) | 7.94 | 8.18 | (8.3) | 41.28 | 41.78 | 21.74 | −21.54 | −22.86 | 0.73 |
| 22.1350 | 0.510 | 0.52 ± 0.12 | −0.32 ± 0.07 | 0.54 | 8.92 | 8.80 | 8.60 | 7.72 | 7.90 | 8.16 | 40.60 | 40.95 | 22.29 | −19.88 | −20.67 | .... |
| 22.1528 | 0.665 | 0.73 ± 0.06 | −0.06 ± 0.04 | 0.74 | 9.08 | 8.60 | (8.3) | 7.94 | 8.10 | (8.3) | 41.07 | 41.51 | 22.17 | −20.54 | −21.17 | .... |

Notes: (a) (8.3) in brackets indicates that the nominal turn-around point was reached, so this is not a real "limit".
(b) Line luminosity in units of erg s$^{-1}$.
(c) Half light radius in kpc
(d) Limit only in H$\beta$ so the error bars in $R_{23}$ are highly asymmetrical and are given as a range.

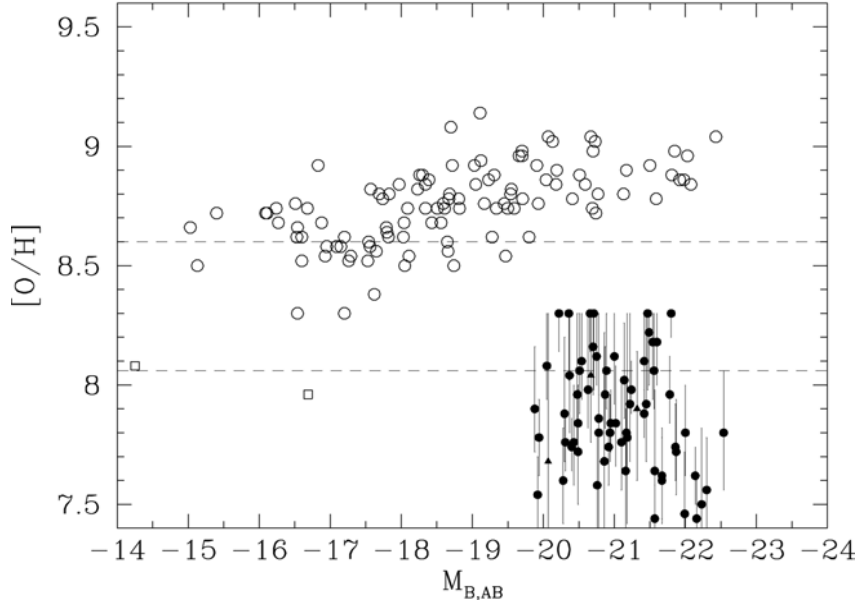

*Fig 10 (b) As for (a) except that it is assumed that all CFRS galaxies are on the low metallicity branch of the $R_{23}$-[O/H] degeneracy. This would give most galaxies extremely low metallicities, which we regard as unlikely.*

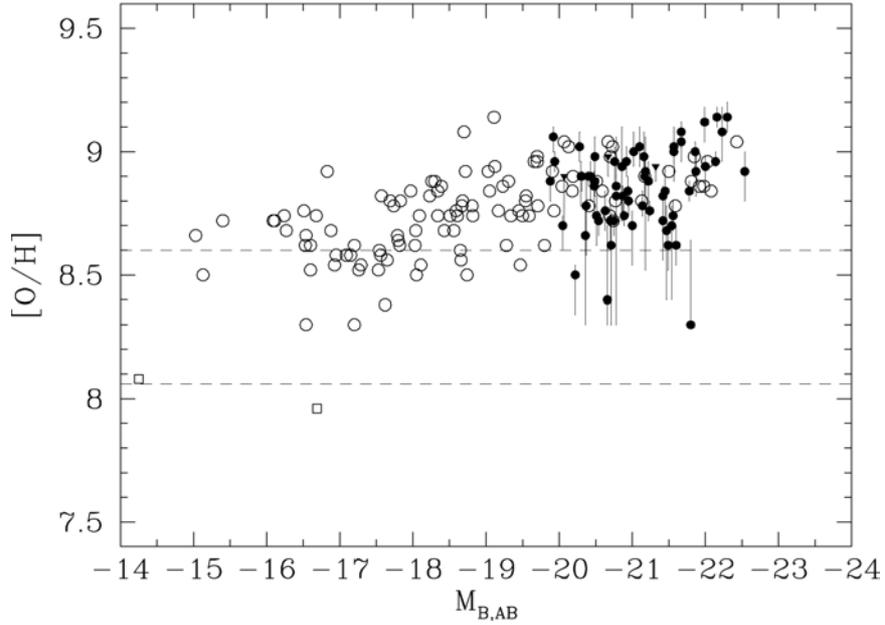

*Fig 10 (c) As for (a) except that all the CFRS galaxies were assumed to be completely unreddened. This substantially reduces the fraction of galaxies with low metallicities [O/H] < 8.6. Since lower metallicity galaxies would be expected to have lower reddening, the true situation may lie between the extremes of (a) and (c)*

*3.5 Comparison with Hammer et al. (2001)*

Hammer et al. (2001) have recently published their study of a sample of 14 CFRS galaxies. Their conclusion was that the metallicity of the gas in these galaxies appeared to be significantly lower (0.25 to 0.4 $Z_\odot$) than the dominant older stellar population (solar to oversolar), leading them to hypothesize that these objects were forming their bulges at a late epoch out of relatively fresh infalling material. While smaller, the Hammer et al. (2001) sample should be directly comparable to the sample here, with similar redshift limits (0.45 < z < 0.8) and the requirement for significant flux in [OII] $\lambda 3727$ (rest equivalent width > 15 $\approx$). Their size/compactness criterion of $r_{0.5}$ < 5 kpc (for $q_0 = 0.5$ and $H_0 = 50$ kms$^{-1}$Mpc$^{-1}$) would correspond to $r_{0.5}$ <

4.5 kpc with our assumed cosmology and it may be seen from Fig. 13 that this range covers 2/3 of our own sample (which had no size selection per se). Indeed 4 of their objects appear in our current sample.

Although they adopted a less simplistic treatment of both extinction and stellar absorption lines, the differences between our conclusions stem mostly from the different treatment of the $R_{23}$ degeneracy. In deriving their metallicities, Hammer et al. appear to have simply averaged the values obtained from the high and low solutions of $R_{23}$ which inevitably leads to values close to the point of reversal at [O/H] ~ 8.4. It is clear from their Table 6 that their ì high solutionsî allow quite high metallicities in many of their galaxies. For example, about half their sample has ì high solutionsî with [O/H] ≥ 8.8. Hammer et al. (2001) emphasized the difference in metallicities between the gas and the solar or above-solar metallicities that they inferred for the dominant stellar populations in their galaxies. While we have not attempted to estimate the latter, solar metallicities for the stars in many of our galaxies would be quite consistent with our broadband colours (see Fig. 16), and with our gas metallicity estimates.

## 4. TRENDS WITH METALLICITY

### 4.1 Properties that do not appear to correlate with [O/H]

We have searched for relations between [O/H] and several properties that we might expect to correlate with it ñ specifically with galaxy morphology and size and with Hβ strength (equivalent width and luminosity). In all of these properties we have not identified a convincing trend with metallicity in our data.

On Fig. 11 we show plots of [O/H] vs. both Hβ luminosity, which should be a crude measure of overall star-formation rate, and Hβ equivalent width, which will be a broad measure of specific star-formation rate. No strong trends are visible between [O/H] and either measure of Hβ strength. However, we were unable to individually correct the Hβ measurement for either underlying stellar absorption or reddening which will degrade its usefulness as a star-formation indicator.

High-resolution F814W HST images are only available for 26/66 (40%) of the sample. The areas 10x10 arcsec$^2$ surrounding each galaxy are shown in Fig. 12. An observed *I*-band half-light radius $r_{0.5}$ (i.e. rest-frame 4400-5500 ≈) has been measured from these images as in Lilly et al. (1998) and this is plotted as a function of the nominal metallicity from the present study in the left hand panel of Fig. 13, together with the equivalent data (measured in rest-frame *B*) for the NFGS sample. Half-light radius is a combination of both the size and the compactness of the galaxy in question.

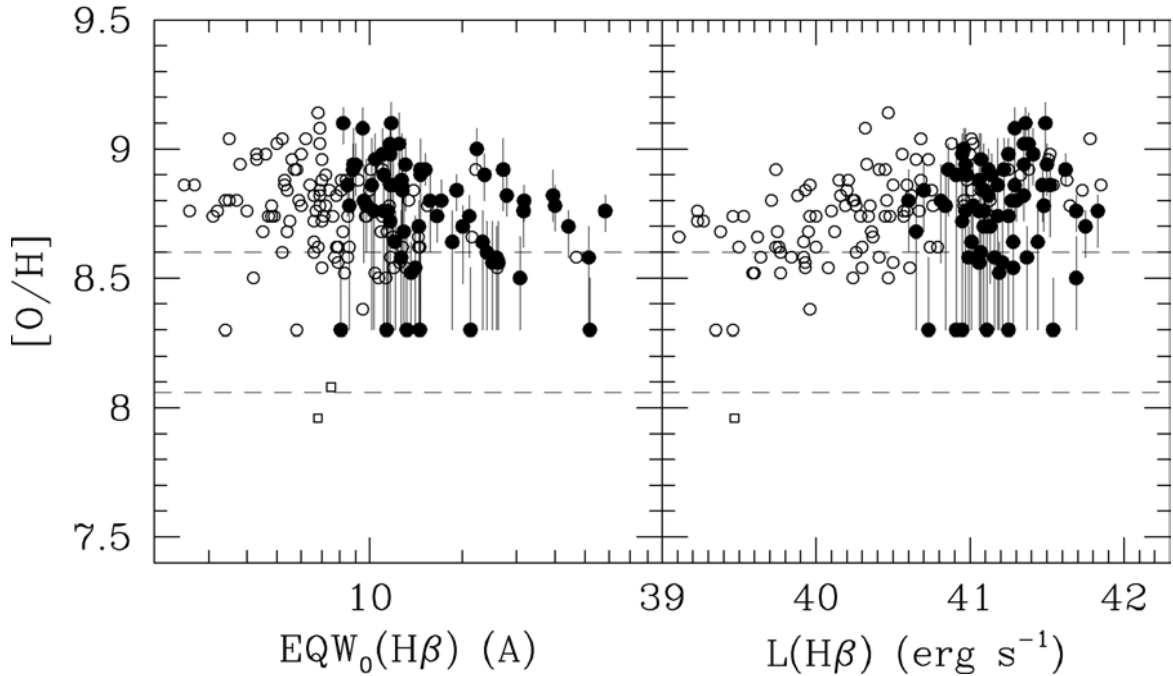

*Fig. 11. [O/H] estimated from $R_{23}$, for the 0.47 < z < 0.92 CFRS galaxies and the NFGS sample (symbols as in Fig 10) as a function of rest-frame Hβ equivalent width (left panel) and Hβ luminosity (right panel), where both measures of Hβ have been corrected for stellar absorption as described in the text. There are no trends in [O/H] with Hβ strength.*

It is clear on Fig. 13 that there are no strong correlations within the CFRS sample between [O/H] and $r_{0.5}$, an impression that may be confirmed by eye on Fig. 12. The higher metallicity CFRS galaxies, [O/H] > 8.6, have half-light radii that are very similar to the range exhibited by the local sample and extend down to the smallest galaxies in our study, with $r_{0.5}$ < 2 kpc. On the other hand, the limited number of apparently lower metallicity CFRS galaxies for which HST images available (noting that their errorbars extend in most cases out of the "turn-around" region), have sizes ($r_{0.5}$ 7 3 kpc) that are relatively large, and certainly larger than the majority of low metallicity galaxies locally, which have $r_{0.5}$ < 3 kpc (Fig. 13). It is also apparent that the four ìlow metallicityî and two ìmedium metallicityî galaxies at the bottom of Fig. 12 do not have unusual morphologies compared with the others, and (as also indicated quantitatively on Fig. 13) they are not the most compact sources in the present sample either. On the other hand, the most regular galaxies morphologically are found in the upper, higher metallicity, part of the figure.

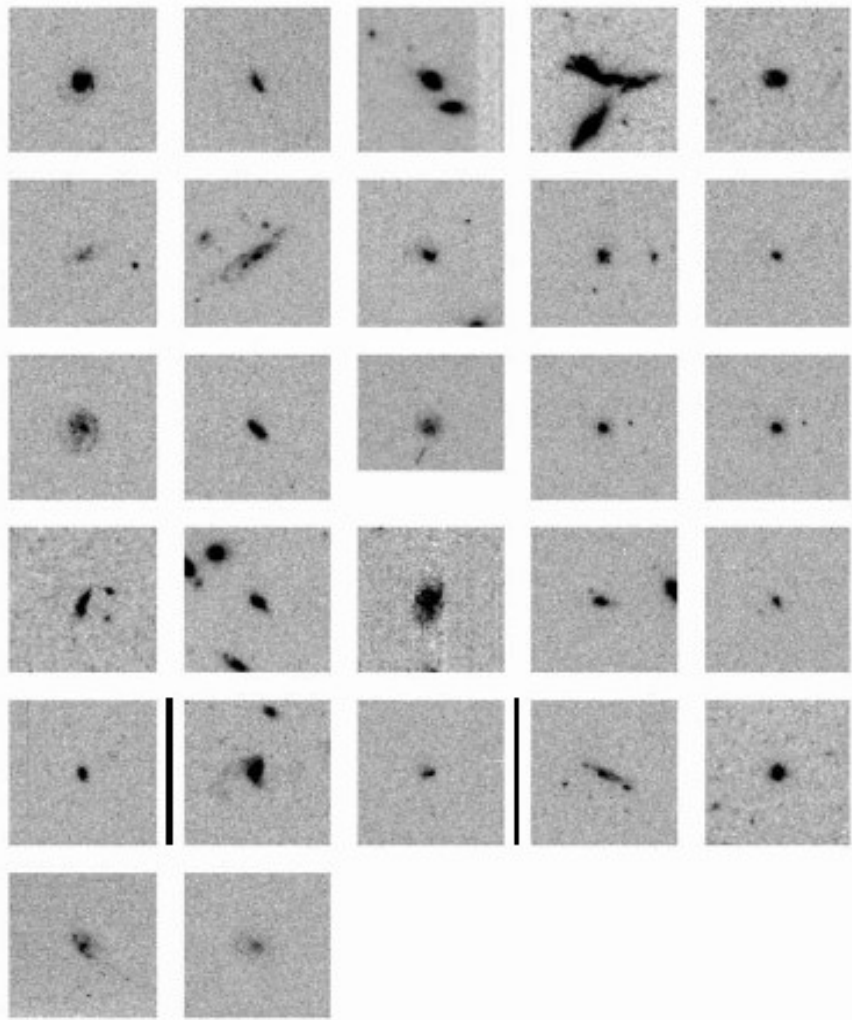

*Fig. 12. HST F814W images of galaxies in the sample arrange in order of decreasing nominal metallicity (left to right, top to bottom). The vertical bars indicate nominal [O/H] ~ 8.7 and ~ 8.6 respectively (c.f. Figs. 15).*

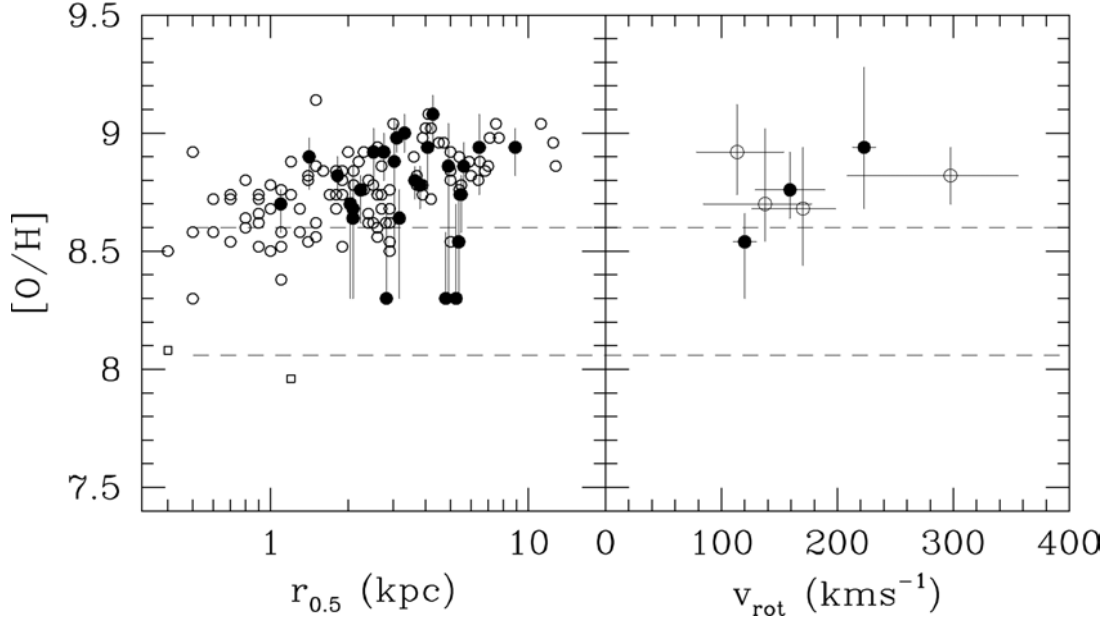

*Fig. 13. (left) [O/H] estimated from $R_{23}$, for the 0.47 < z < 0.92 CFRS galaxies and the NFGS sample (symbols as in Figs. 10 and 11) as a function of half-light radius $r_{0.5}$ for the 26 galaxies with HST F814W images available. As with luminosity (see Figs. 10 and 17) the higher metallicity galaxies overlap the region of this diagram occupied by the NFGS sample, while the lower metallicity CFRS galaxies are considerably larger than present-day galaxies of the same metallicities. (right) [O/H] for the handful of galaxies with kinematic data: Solid symbols represent true rotation velocities from Barden et al (2003) while the open symbols have been derived from the velocity dispersion data of Mallen-Ornelas et al (1999) by multiplying by 1.7. With the limited data set no strong trend is seen although it can be seen that the three rotation velocities of Barden et al do increase monotonically with metallicity.*

Finally, there are a handful of these galaxies that have kinematic data in the emission lines. Three galaxies have recently published Hα rotation curves (Barden et al 2003). For another four, Mallen-Ornelas et al (1999) measured integrated velocity dispersions, from which we have computed an effective $v_{rot}$ by multiplying by 1.7, derived assuming a square line profile, so these rotation measures may be quite uncertain. The right hand panel in Fig. 13 shows the metallicities that we have derived for these seven galaxies as a function of their estimated $v_{rot}$. There is again no strong trend - the best that can be said is that there may be a monotonic relation with the three rotation velocities from Barden at al data. In this context, it is worth noting that all four of our galaxies observed by Mallen-Ornelas et al had σ above the median of their sample of 36 galaxies.

*4.2 Correlations between[O/H] and broad-band colors*

K-band photometry is available for 54/66 (82%) of the galaxies in our statistically complete sample. At these redshifts the observed K-band is close to the rest-frame J-band. Fig. 14 shows the observed $(V-I)_{AB}$ vs. $(I-K)_{AB}$ plane for the sample galaxies with available K photometry together with the tracks for galaxies of different Hubble types as a $f(z)$ for 0.47 < z < 0.92. These have been derived from the Coleman et al. (1980) spectra as in Lilly et al. (1995). They have not been de-reddened and can be compared directly with the observed colors of the CFRS sample. The ìhigh metallicityî sub-sample cluster around the expected colors of late type galaxies, while the lower metallicity sub-samples extend down to significantly bluer $(I-K)_{AB}$ colours whilst exhibiting a similar range of $(V-I)_{AB}$.

The strong correlation that is implied on Fig. 14 between metallicity and the rest-frame colours is shown directly in Fig. 15. A Spearman rank test indicates that the correlation with $(B-J)_{AB,0}$ is significant at 5σ and is the strongest found between our [O/H] measurements and any other property of the galaxies. The correlation with $(U-V)_{AB,0}$ is weaker but still significant at 4σ. The colour and metallicity data are completely independent from an observational point of view.

In order to interpret these correlations, Fig. 16 shows the rest-frame $(U-V)_{AB}$ vs. $(B-J)_{AB}$ plane together with several theoretical stellar population models derived from the 2001 Bruzual and Charlot models (Bruzual & Charlot, 1993, updated on http://umpu.cida.ve/~bruzual/bcXXI.html). The effect of a $E_{B-V} = 0.3$ ($A_V = 1$) extinction is shown. Given the difficulties of transforming between different color systems, most attention should be paid to differential effects. The tracks on Fig. 16 show models with a single initial burst of star-formation and with a constant star-formation rate, both computed at both solar metallicity (Z = 0.02) and at quarter solar metallicity (Z = 0.005). The initial mass function in all cases is a broken Salpeter power-law with $x = 0.3$ for 0.1 to 0.5 $M_\odot$ and $x = 1.3$ for 0.5 to 100 $M_\odot$. The symbols along each track (starting from the upper right) reflect ages of 14, 8, 4 Gyr and thereafter successively halving in age.

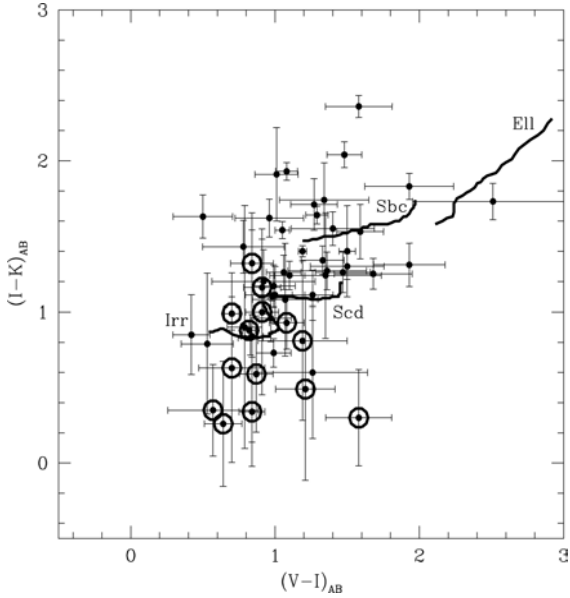

*Fig. 14. The observed $(V-I)_{AB}$ vs. $(I-K)_{AB}$ plane for the 54 CFRS galaxies with K photometry. Galaxies with nominal [O/H] < 8.7 (computed as in Fig. 10a) are highlighted by surrounding circles and systematically show bluer (I-K) colors, and to a lesser extent (V-I) colors, than the remainder. The tracks show the colors as a function of redshift expected for standard galaxy spectral energy distributions computed from Coleman et al (1980) from z = 0.47 (left hand end of track) to z = 0.92 (right hand end).*

In the following discussion, we explore possible causes of the color-[O/H] correlation and explore for each of them the effect on the *J*-band mass-to-light ratio in an effort to understand the masses of the lower metallicity galaxies.

If the dominant stellar population reflects the [O/H] metallicity of the gas sampled by $R_{23}$, a small change in colour would be expected from direct opacity effects in the stellar atmospheres. This is quite a modest effect ñ about 0.3 magnitudes in $(B-J)_{AB}$ and very little in $(U-V)_{AB}$. The *J*-band mass-to-light in the low metallicity model is about 0.3 magnitudes lower. The offset exhibited by the bulk of the lower metallicity galaxies (highlighted in the Fig. 16) relative to the remainder is a little larger than this ñ of order 0.6 mag on average in (*B-J*) and about 0.3 mag in (*U-V*), leading to a suggested additional effect of about 0.3 mag in both colours. Inspection of the figure shows that this could be due to a number of plausible physical effects.

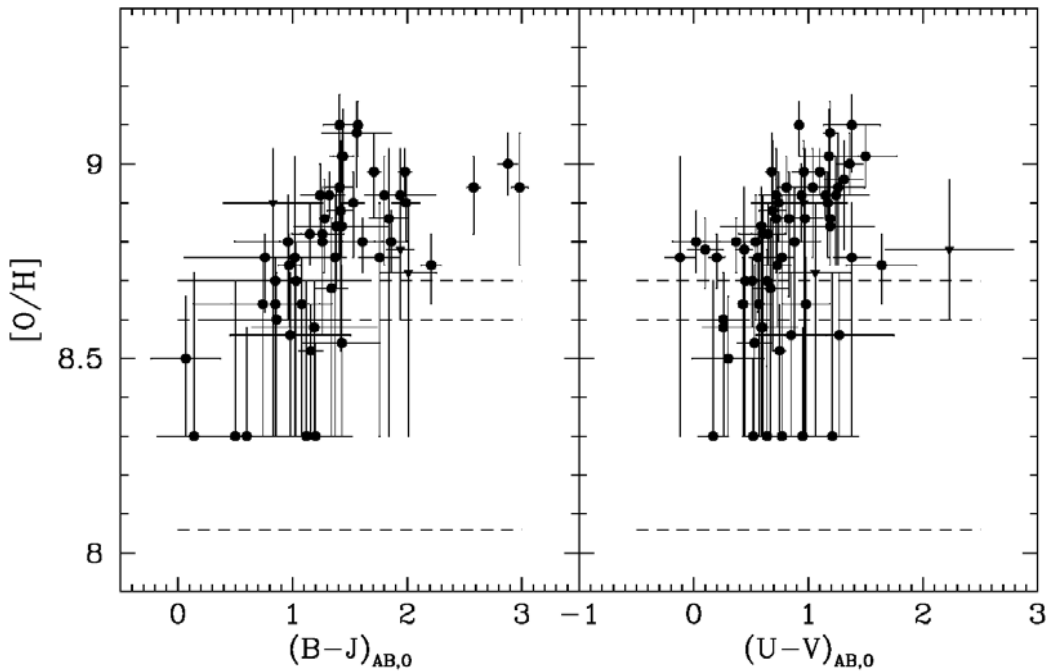

*Fig. 15. Rest-frame $(B-J)_{AB,0}$ and $(U-V)_{AB,0}$ colors of CFRS galaxies plotted against nominal [O/H] (as computed for $E_{B-V}$ = 0.3 as in Fig. 10a). The horizontal dashed lines show the turn-around region where the metallicity is very sensitive to $R_{23}$ and the reddening. There is a good correlation with (B-J) in the sense that the lowest metallicity galaxies have systematically bluer (B-J) colors.*

*(a) Reddening by dust*

Greater reddening of the continuum in the higher metallicity galaxies would have a bigger effect in (*B-J*) than in (*U-V*) and a differential extinction of only $\Delta E_{B-V} \sim 0.10$ mag would fully account for the remaining offset in (*B-J*) and (*U-V*). This is 1/3 of the mean reddening applied to the spectra (see Fig. 9) and is therefore entirely plausible if the lower metallicity galaxies are only weakly reddened. It should be noted that variations in the dust extinction at this level would have a negligible effect on the mass-to-light ratio in the *J*-band.

*(b) Overall younger ages*

An overall younger age for the stellar population would also produce the required effect. Based on the continuous star-formation models, a decrease in age of about a factor of four would be required. In the models on the figure ñ an offset of about 0.25 mag in (*U-B*) and 0.35 in (*B-J*) is produced by changing the age of the constant burst model by a factor of four, i.e., from 8 Gyr at 1 $Z_\odot$ to 2 Gyr at $0.25 Z_\odot$. The associated change in mass-to-light in the *J*-band is

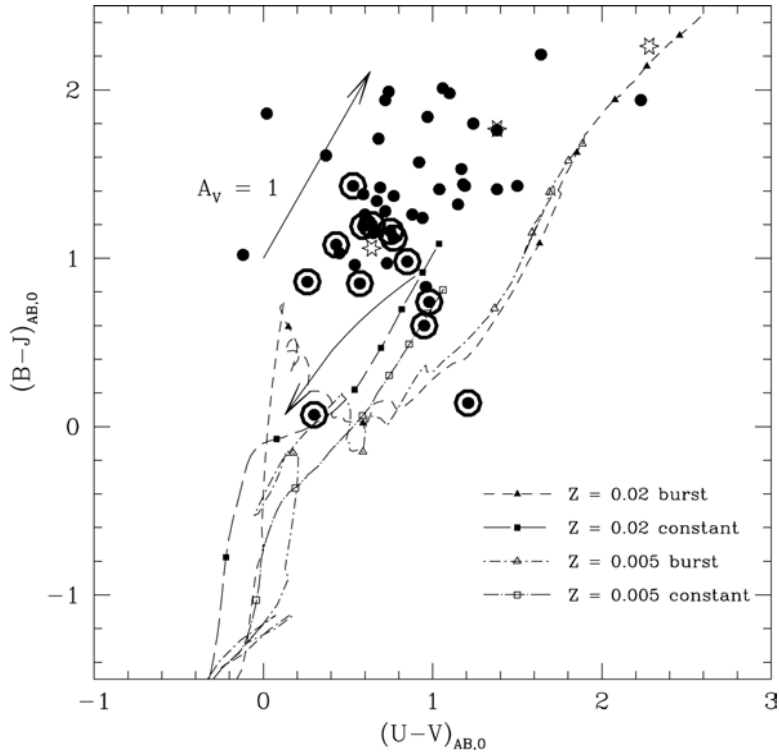

*Fig. 16. The rest-frame (B-J)$_{AB}$ vs. (U-V)$_{AB}$ two-color diagram showing the CFRS galaxies with available K photometry. Galaxies with nominal [O/H] < 8.7 (computed as in Fig. 10a) are highlighted as in Fig. 14. The large open stars show, in order of decreasing color, the colors of Ell, Sbc and Irr galaxies from Coleman et al. (1980). The two sets of lines show stellar population models from Bruzual and Charlot (2001) computed for single burst (dashed lines) and constant star-formation (continuous lines) models. The points along each track show ages (from the top right) of 14 Gyr, 8 Gyr, 4 Gyr and thereafter halving in age. In each case, the models are computed for metallicities of Z = 0.02 and Z = 0.005. The straight arrow is a reddening vector of E$_{B-V}$ = 0.3 while the curved arrow shows the effect of adding a 0.01 Gyr population to the 8 Gyr constant star-formation model. The high redshift CFRS galaxies cluster around the Irr galaxy extending up towards the Sbc galaxy. The lower metallicity galaxies (highlighted) are concentrated at the blue end of this distribution and show a significant offset in their mean colors compared with the rest of the sample. This offset plausibly arises from a combination of photospheric effects and lower reddening and/or younger ages in lower metallicity galaxies. None of these will produce a large variation in the J-band mass-to-light ratio (see text for discussion).*

0.85 magnitude.

doubt be constructed to match the limited data, the

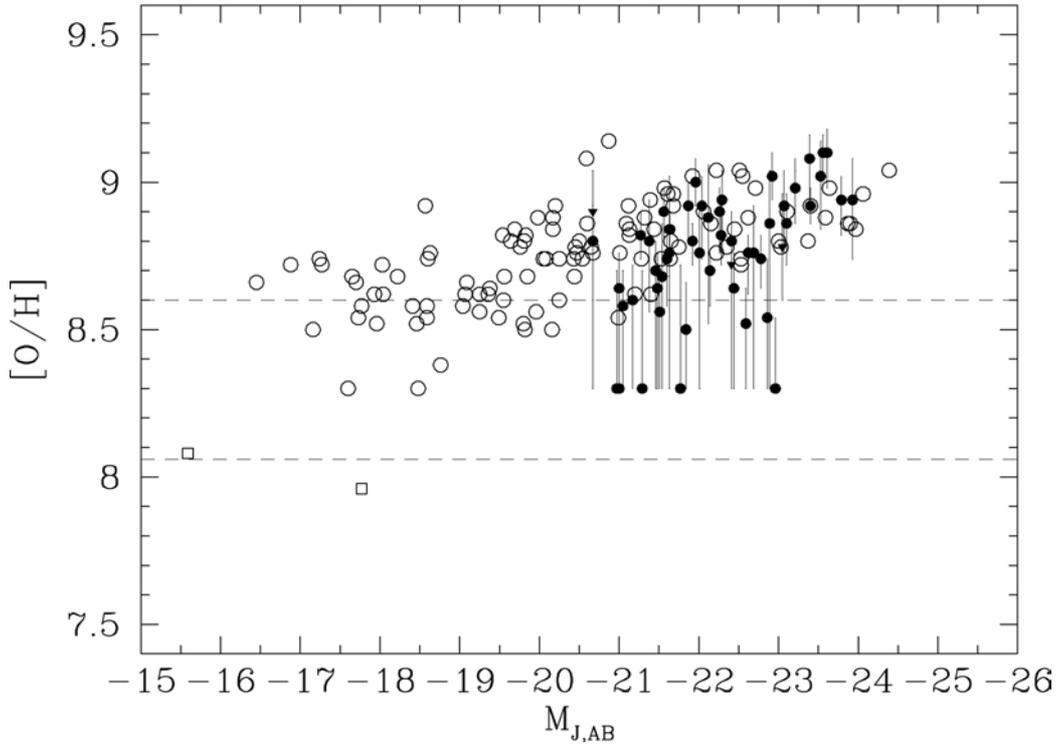

*Fig. 17. The metallicity of CFRS galaxies, derived for a uniform $E_{B-V} = 0.3$, as a function of J-band absolute magnitude $M_{J,AB}$ for the 54 galaxies with K photometry. The higher metallicity CFRS galaxies overlap well the region of the diagram occupied by NFGS of the same luminosity, whereas the lower metallicity CFRS galaxies occupy a region displaced by several magnitudes in luminosity and must have migrated by the present epoch.*

*(c) Addition of a recent starburst*

Addition of a burst of star-formation (assuming the star-formation averaged over the last $10^7$ years has been three times the previous constant level) is sufficient to produce a change in both colours of 0.3 mag, but changes the *J*-band mass-to-light ratio hardly at all (0.06 mag).

With our current data, it is clear that we cannot convincingly differentiate between these alternatives, although we note that the apparent colours of the most extreme galaxies with apparent rest-frame $(B-J)_{AB} < 0.5$ would require a stellar population that is less than 4 Gyr old on average.

The point of this discussion is to note that none of these simple explanations for the offset in $(B-J)$ produce a mass-to-light change in *J* that is larger than 1.2 magnitudes relative to the remainder of the sample. Constant star-formation models may additionally fade in J by about 1 magnitude to the present epoch if star-formation ceases. The cumulative effect is that it is unlikely that the ì low metallicityî CFRS galaxies identified in this paper can fade by more than 2.5 magnitudes to the present epoch. While contrived models could no important point is that the color changes that we have observed to be correlated with the metallicities are most easily explained by models that involve only modest changes in the *J*-band *M/L* ratio.

5. DISCUSSION

*5.1 Evolutionary descendents of individual galaxies*

Fig. 10a and Fig. 13a compared the nominal metallicities of the CFRS galaxies as functions of $M_B$ and half-light radius with the NFGS local sample. Fig. 17 shows the equivalent to Fig. 10a for $M_{J,AB}$. One consequence of the blue $(B-J)_{0,AB}$ colors for the lower metallicity galaxies is that the absence of high luminosity low metallicity galaxies noted above in $M_{B,AB}$ becomes more prominent in $M_{J,AB}$.

Fig. 13a and Fig. 17 both give an impression of a dominant upper envelope matching the high metallicities of the NFGS sample and extending up to the highest luminosities and largest sizes, with the addition of a significant population or tail extending to lower metallicity galaxies. These are about 4 magnitudes brighter in $M_J$ and (from a rather small sample) a factor of 3 larger in size than objects with comparable metallicities in the NFGS sample. As noted above they comprise between 5-25% of the sample depending on the reddening correction.

It was argued above that the simplest explanations for the colors and color differences observed in the galaxies as a $f(O/H)$ suggest only modest changes in M/L ratio. Furthermore, the sizes of the lowest metallicity galaxies also suggest that they will not evolve into the smaller low metallicity galaxies locally. Thus we conclude that the most likely explanation of the < 25% of the sample exhibiting lower metallicities ([O/H] < 8.6) is that they will evolve into massive high metallicity galaxies similar to the remainder of the sample rather than remaining at low metallicities. In essence, the evolution in Figs. 13a and 17 will be primarily vertical and not horizontal.

This picture thus supports the so-called "downsizing" view of galaxy evolution in which the manifestations of activity, including strong emission lines, blue colours, disturbed morphologies and, with this work, moderate metallicities, are found in progressively more massive galaxies at higher redshifts, and argues against the idea that the luminous blue galaxies at $z > 0.5$ are highly brightened dwarf galaxies.

What are these galaxies which appear to have evolved significantly in metallicity over the last half of the age of the Universe? We may speculate that they are similar to the remainder but have, for some reason, progressed less far along their evolutionary paths in the first half of the life of the Universe. As noted above, the fraction of galaxies involved is quite uncertain (5ñ25%) on account of the possibly over-estimated reddening correction for the lower metallicity galaxies. The metallicity ì shortfallî for all but a handful of extreme galaxies could be quite modest.

Over the last ten years, the idea that galaxies similar to present-day dwarf galaxies (albeit over-luminous by 2-3 magnitudes) were playing a major role in the galaxy population at intermediate redshifts ($z < 0.5$) and flux densities (B < 24) has been widely discussed (e.g., Cowie et al. 1991). It was imagined that such galaxies could either fade into the dwarfs seen today or merge into more massive galaxies. The arguments in favour of this picture arose principally from the apparently high number density of such galaxies in the sub-L* regime where the galaxy luminosity function is relatively flat. In addition, several studies showed relatively low dynamical velocities in samples of galaxies selected in various ways (e.g., Guzman et al. 1997, 1998, Simard & Pritchet, 1998, Forbes et al. 1996, Phillips et al. 1997, Mallen-Ornelas et al. 1999). Furthermore, the association was supported by the observation of relatively small $r_{0.5}$ ~ a few kpc, irregular morphologies, blue colours and strong emission lines and other properties frequently found in ìdwarfî galaxies. Finally, there has undoubtedly been a guilt by association with the myriad of very small galaxies seen at much fainter levels in the HDF, but which contribute little to the overall surface brightness of the sky and thus to the overall star-formation history.

The metallicities of the gas in the CFRS galaxies do not support the idea of fading dwarfs dominating the galaxy population, at least at these L* luminosities. In most galaxies the gas is already close to solar in metallicity and thus significantly higher than found in present-day dwarf galaxies. We can be confident that most of the stars being produced at $0.5 < z < 1.0$ end up in massive galaxies today. This is reassuring ñ an objection to the fading dwarf picture has always been that this would place too many stars in systems where few stars reside today. This aspect was emphasized by Ferguson and Babul (1998) who showed that a $z < 1$ dwarf-dominated model would produce too many faded dwarfs in deep HST images unless they exhibited unusually top-heavy initial mass functions.

What has happened to the evidence in favour of dwarfs? First, it should be appreciated that the changes in the metric associated with recent developments in favour of an accelerating Universe act to reduce the arguments in favour of dwarfs (see e.g., discussion in Lilly et al. 1991). Most notably, the inferred comoving number density of galaxies is reduced by a factor of 2.5 at $z \sim 0.75$ (for the concordance model considered here relative to an $\Omega = 1$ Einstein-de Sitter cosmology). In addition, galaxies are inferred to be larger, by 25%, and more luminous, by 0.5 mag, in the new cosmology. Regarding the kinematic evidence, it has always been understood that measures of velocity dispersions in compact spatially unresolved systems may lead to underestimated masses. Barton and van Zee (2001) give an extended discussion of this topic. It may be that in many cases we are seeing internal disk-bulge evolutionary processes (e.g., Carollo 1999, Carollo et al. 2001).

*5.2 Comparison with Lyman break galaxies*

Although the bulk of our objects have close to solar metallicities, there are some galaxies that apparently have lower metallicities, [O/H] ~ 8.4, and it is interesting to speculate as to the relationship of these objects to the handful of Lyman Break Galaxies (LBG) at $z \sim 3$ for which the $R_{23}$ method has been applied (Pettini et al. 2001). For all of these, similar metallicities significantly below [O/H] ~ 8.9 have been derived for the gas (and for the stars).

Returning to Fig. 6, it is clear that, although the metallicities of the LBG and the least metal-rich CFRS galaxies are similar, they occupy quite different areas of the $R_{23}$-$O_{32}$ plane, with the LBG having higher ionizations with $O_{32} >> 1$. In our own sample, only one galaxy (CFRS 22.0919) has a similar spectrum. This emphasizes that the LBG experience different physical conditions than the most luminous galaxies at $z < 1$ even when the metallicities are similar.

*5.3 The mean metallicity of star-forming gas as f(epoch)*

One quantity of interest is to try to estimate the global mean metallicity of gas out of which stars are being made in the Universe as a function of cosmic epoch. The average [O/H] abundance of all 66 $0.47 < z < 0.92$ CFRS galaxies, weighted by H$\beta$ luminosity as a measure of star-formation rate, is [O/H] = 8.81 with a statistical uncertainty of only about 0.02. The same quantity for the NFGS galaxies with $M_{J,AB} < -20$ (see Fig. 17) is [O/H] = $8.89 \pm 0.01$.

The difference of dex $0.08 \pm 0.02$ in the means is quite small, and is comparable to the probable systematic uncertainties, including reddening. The global uncertainty in translating $R_{23}$ to [O/H] probably has a small effect since the spectra of both high and low redshift samples are similar (Fig. 6) and have been treated in the same way, so the uncertainties in the calibration presumably affect both samples equally. However, a significant uncertainty at the present time involves the reddening in the CFRS sample, which was assumed to be 0.30 magnitudes in $E_{B-V}$. An error in the average extinction of 0.15 mag (c.f. Fig. 9) would translate to an uncertainty in the mean [O/H] of 0.05 dex for this sample. Finally, the treatment of the galaxies in the ìreversal zoneî is at present rather crude and as discussed above, many of the individual galaxies in this region could have metallicities between 8.05 < [O/H] < 8.6. We estimate that the treatment of these galaxies introduces an additional uncertainty of order 0.03 dex in the mean metallicity.

Thus our best estimate for the change in metallicity of the star-forming gas in luminous, and as argued above, reasonably massive, galaxies is $\Delta$[O/H] ~ 0.08 ± 0.06 over the 50% of cosmic time that has elapsed between $z$ ~ 0.75 and the present.

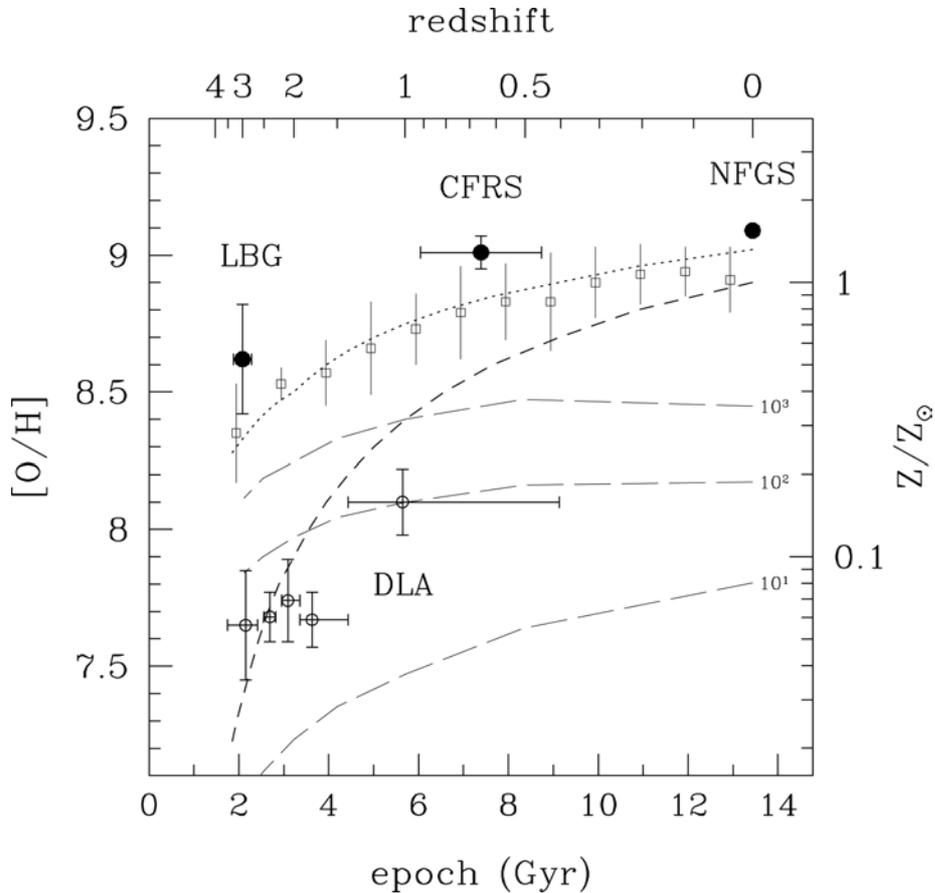

*Fig. 18. Estimates of metallicity over cosmic time. The heavy solid symbols represent [O/H] metallicities (left hand axis) at three cosmic epochs derived in a self-consistent way from $R_{23}$ using the data for the NFGS in JFFC, this paper and Pettini et al. (2001). The first two are derived from large samples weighted by H$\beta$-luminosity. The light open circles are DLA absorption systems and are based on [Fe/H] metallicities (right hand axis) in the column-density weighted analysis of Kulkarni & Fall (2002). The evenly spaced open squares are the [Fe/H] age-metallicity relation for the Galactic disk from Twarog (1980). The two axes are linked assuming [O/H]$_\odot$ = 8.69 (Allende Prieto et al 2001). It should be noted that the older value of [O/H]$_\odot$ = 8.9 results in a better overlap of the Twarog points and the CFRS and NFGS data. The lines represent various theoretical models: the global model from Pei et al. (1999) (short dashed line); the collisional star-burst model from Somerville et al. (1999) (dotted line); and three cuts of overdensity from the numerical simulations of Cen and Ostriker (1999) (long dashed lines).*

Of course, some star-formation will be occurring in galaxies below $M_{J,AB}$ < -20. The CFRS samples of order 50% of the global star-formation rate at 0.5 < $z$ < 1.0, based on optical emission line strength (i.e. excluding heavily obscured galaxies). Weighting the NGDS metallicities by the Gallego et al. (1996) H$\alpha$ luminosity function we find that the mean [O/H] of the galaxies contributing the top 50% of the H$\alpha$ luminosity function is only $\Delta$[O/H] = 0.08 higher than that of the integrated population. It is our view that it is reasonable to suppose that the $\Delta$[O/H] ~ 0.08 ± 0.06 derived above for luminous galaxies in the CFRS and NFGS will be valid also for star-formation in the Universe as a whole.

*5.4 Comparison with models for metal enrichment history*

Fig. 18 shows the average (H$\beta$-weighted) $R_{23}$-derived [O/H] metallicities for our sample and the local NFGS sample with $M_{J,AB}$ < -20 (as derived above in Section 4.4b) and an unweighted average of the 5 LBG of Pettini et al. (2001). The latter are shown assuming that they are on either the ìhighî and ìlowî metallicity branches, although the values of $R_{23}$ are high enough that the difference that this makes is much smaller than would be the case for the present sample.

The LBG are ultraviolet selected and so might conceivably be biased to lower metallicities relative to the other two samples (cf. Contini et al. 2002). Nevertheless, within the error bars, the gradual increase with epoch shown in the figure is, as argued above, likely representative of the change in mean metallicity of stars forming in the Universe.

In comparison, Fig. 18 also shows the mean (column density weighted) [Fe/H] metallicity of neutral Hydrogen in DLA absorption systems derived from Zn abundances by Kulkarni and Fall (2002), and the [Fe/H] age-metallicity relation from the Galactic disk from Twarog (1980, see the discussion in Garnett & Kobulnicky 2000) using the scale on the right. The right and left hand axes are linked using [O/H]$_\odot$ = 8.69 (Allende Prieto et al 2001).

Fig. 18 emphasizes the substantial difference between DLA gas metallicity and the metallicity of the gas out of which stars are actually being made. This is no doubt because the emission line measurements are sampling the highest density and presumably most enriched parts of the Universe at any epoch. On the other hand, the apparent consistency with the disk age-metallicity relation is impressive, and would be more so if the older [O/H] ~ 8.9 solar value was retained.

Several recent models for the evolution of metallicity in the Universe are shown: the global model for neutral gas from Pei et al. (1999), the numerical simulations of Cen and Ostriker (1999) at three different overdensity thresholds, and the semi-analytic ìcollisional-starburstî model of Somerville et al. (1999), again for neutral gas. All three models are broadly consistent with estimates of the global star-formation rate with cosmic epoch. Any comparison of these models with the data needs to be approached with considerable care, not least because of the density effects. As an example, there is no reason why the Pei et al. (1999) model for the global metallicity of neutral gas should reproduce the metallicity of the star-forming gas.

Overall, the Somerville et al. (1999) ìcold gasî curve does quite well in reproducing the overall trend of star-forming metallicity with redshift, even though it does a relatively poor job at matching the DLA data (it cannot fit both). Looking at relative effects, both the Somerville et al (1999) and the Pei et al. (1999) model have larger changes over the last half of the age of the Universe (0.2 and 0.3 dex respectively) than are seen in our data although the difference with the former at least is not very significant. Interestingly, many models for the evolution of the solar neighbourhood (e.g., Prantzos & Silk 1998) also predict a steeper rise with epoch than is seen in the age-metallicity data.

The models which best match the weak evolution of average metallicity (but which are offset in actual [O/H] value] are the highest density regions in the hydro-dynamical simulations of Cen and Ostriker (1999). As emphasized by them, the development of metallicity is strongly density dependent.

# 6. CONCLUSIONS

The metallicity of the gas associated with star-formation in a statistically complete H$\beta$-defined sample of 66 CFRS galaxies with 0.47 < $z$ < 0.92 (i.e. a look-back of about half the age of the Universe) have been estimated using the $R_{23}$ parameter. The sample is defined to have $I_{AB}$ < 22.5, $f_{3727}$ > $10^{-16}$ erg s$^{-1}$ cm$^{-2}$ and rest-frame absorption-corrected EQW$_0$(H$\beta$) > 8 $\approx$. The sample is representative of the galaxies producing 90% of the [OII] $\lambda$3727 emission from the CFRS sample at this epoch, which itself is likely representative of about a half of all the non-heavily obscured star-formation in the Universe at these redshifts. From these data, four main conclusions have been derived:

1. With very few exceptions, CFRS star-forming galaxies (i.e., $M_{B,AB}$ < -20, EW(H$\beta$) > 8 $\approx$) at 0.4 < $z$ < 0.9 have line ratios in the bright lines of [OII] $\lambda$3727, [OIII] $\lambda$5007 and H$\beta$ that are very similar to those seen across a large sample of local galaxies (-14 < $M_{B,AB}$ < -23). This is particularly true if the average reddening in the CFRS galaxies is higher than in the local comparison sample, as would be expected given the different luminosity ranges.

2. Making a reasonable assumption about their average reddening ($A_V \sim 1$), the bulk of CFRS galaxies have metallicities [O/H] > 8.6 that are similar to those seen in local galaxies of similar luminosities. However, a significant fraction appear to have lower [O/H] abundances that are seen in the present Universe in galaxies with luminosities only a factor of 10 or more lower. The fraction of lower metallicity galaxies could be as high as about 25% with our reddening assumption, but could be reduced to only about 5% if the reddening is much less than assumed (especially for the lower metallicity galaxies).

3. The lower metallicity galaxies, inferred to have [O/H] ~ 8.4 (with considerable uncertainty) have bluer colours in rest (*U-V*) and, especially, in rest (*B-J*) than the higher metallicity ones. This may be naturally explained in terms of photospheric effects, lower extinction and/or younger stellar ages, none of which would produce large changes in the *J*-band M/L ratio of more than 1.2 magnitude. For the subset with HST images, the galaxies with the most well developed spiral morphologies are all of high metallicity. However, those with apparently lower metallicities are also quite large, $r_{0.5} > 3$ kpc, larger than many higher metallicity galaxies (although the error bars in [O/H] extend up towards the other objects). The single lower metallicity galaxy with a rotation velocity measurement has $v_{rot} > 100$ kms$^{-1}$. It is thus supposed that these are immature massive galaxies that will increase their metallicity to the present epoch, in a "down-sizing" scenario, rather than highly brightened ìdwarfî galaxies that might have remained at low metallicities to the present-day while fading in luminosity.

4. The overall metallicity in the sample is only slightly lower at half the present age of the Universe than that seen in comparably luminous galaxies today, $\Delta$[O/H] = 0.08 ± 0.06, with the uncertainty dominated by our uncertainties in the reddening of individual galaxies. This modest change is in accord with the local age-metallicity relation in the Galactic disk and is broadly consistent with models for the development of metallicity in the Universe, particularly with those that differentiate between regions of different densities.

The new sample that we have presented significantly extends the study of [O/H] abundances of star-forming to earlier epochs. Broadly speaking, the results show only modest evolutionary effects relative to similar galaxies observed locally. On the other hand, the appearance of some apparently lower metallicity galaxies suggests that we may be approaching an epoch where variations in the evolutionary tracks of galaxies which today may appear quite similar are becoming apparent.

We are grateful to the anonymous referee for his/her very careful reading of the manuscript and for numerous helpful suggestions that have improved the paper. We are grateful to Jarle Brinchmann, Mark Brodwin and Tracy Webb for providing unpublished *K*-data for some of the galaxies in this analysis. Thanks are also due to Marianne Takayima and Tom Geballe for their service observing of CFRS 14.0393 during the Gemini run on Keck. Most of the data in this paper were obtained with the Canada-France-Hawaii Telescope, jointly operated by the National Research Council of Canada, the Centre National de la Recherche Scientifique of France and the University of Hawaii. Some of the data presented here were obtained at the W. M. Keck Observatory, which is operated as a scientific partnership among the California Institute of Technology, the University of California, and the National Aeronautics and Space Administration, and which was made possible by the financial support of the W. M. Keck Foundation. The authors recognize the role of the summit of Mauna Kea in indigenous Hawaiian culture and are grateful to have had the opportunity to conduct scientific observations from it.

REFERENCES

Allende Prieto, Lambert & Asplund 2001, 556, L66

Barden, M., Lehnert, M.D., Tacconi, L., Genzel, R., (1), White, S.D.M., Franceschini, A., 2003, astro-ph/0302392

Barton, E., van Zee, L., 2001, ApJ, 550, 35.

Brinchmann, J., Abraham, R., Schade, D., Tresse, L., Ellis, R., Lilly, S., Le FËvre, O., Glazebrook, K., Hammer, F., Colless, M., Crampton, D., & Broadhurst, T. 1998, ApJ, 499, 112.

Bruzual, G. & Charlot, S., 1993, ApJ, 405, 538.

Carollo, C.M. 1999, ApJ, 523, 566.

Carollo, C.M., Stiavelli, M., de Zeeuw, P.T., Seigar, M., & Dejonghe, H. 2001, ApJ, 546, 216.

Carollo, C.M. & Lilly, S.J. 2001, ApJ, 548, L157 (Paper 1)

Carollo, C.M., Lilly, S.J., & Stockton, A.N., 2002, In ìChemical Enrichment of Intracluster and Intergalactic Mediumî, ASP Conference Proceedings Vol. 253. (eds. R. Fusco-Femiano and F. Matteucci). San Francisco: Astronomical Society of the Pacific, p.167

Cen, R. & Ostriker, J.P., 1999, 519, 109.

Coleman, G.D., Wu, C.C., & Weedman, D.W., 1980, ApJS, 43, 393.

Contini, T., Treyer, M.A., Sullivan M., & Ellis, R.S. 2002, MNRAS, 330, 75.

Cowie, L.L., Songaila, A., & Hu, E.M., 1991, Nature, 354, 460.

Ellison,S.L., Yan, L., Hook, I.M., Pettini, M., Wall, J.V., & Shaver, P. 2001, A&A, 379, 393.

Fall, S.M. & Pei, Y.C. 1993, ApJ, 403, 7.

Ferguson, H.C., Babul, A., 1998, MNRAS, 296, 585.

Forbes, D. A., Phillips, A. C., Koo, D. C., & Illingworth, G. D. 1996, ApJ, 462, 89

Gallego, J., Zamorano, J., Rego, M., & Alonso, O., 1996, ApJ, 459, 43.

Garnett, D.R. & Kobulnicky, H.A., 2000, ApJ, 532, 1192.

Guzm·n, R., Gallego, J., Koo, D. C., Phillips, A. C., Lowenthal, J. D., Faber, S. M., Illingworth, G. D., & Vogt, N. P. 1997, ApJ, 489, 559

Guzm·n, R., Jangren, A., Koo, D. C., Bershady, M. A., & Simard, L. 1998, ApJ, 495, L13

Hammer, F., Crampton, D., Le Fevre, O., Lilly, S.J., 1995, ApJ, 455, 88.

Hammer, F., Gruell, N., Thuan, T.X., Flores, H., & Infante, L. 2001, ApJ, 550, 570

Jansen, R.A., Fabricant, D., Franx, M., & Caldwell, N. 2000, ApJS, 126, 331 (JFFC)

Kewley, L.J. & Dopita, M.A. 2002, astro-ph/0206495.

Kobulnicky, H.A., Kennicutt, R.C., & Pizagno, J.L. 1999, ApJ, 514, 544.

Kobulnicky, H.A. & Zaritsky, D. 1999, ApJ, 511, 118.

Kobulnicky, H.A, & Koo, D.C. 2000, ApJ, 545, 712.

Kulkarni, V.P. & Fall, S.M. 2002, astro-ph/0207661.

Lilly, S.J., Cowie, L.L., & Gardner, J.P 1991, ApJ, 369, 79.

Lilly, S.J., Hammer, F., Le FËvre, O., & Crampton, D. 1995, ApJ, 455, 75.

Lilly, S.J., Le FËvre, O., Hammer, F., & Crampton, D. 1996, ApJ, 460, L1.

Lilly, S.J., Schade. D., Ellis, R., Le Fevre, O., Brinchmann, J., Tresse, L., Abraham, R., Hammer, F., Crampton, D., Colless, M., Glazebrook, K., Mallen-Ornelas, G., Broadhurst, T., 1998, ApJ, 500, 75.

Le Fevre, O., Crampton, D., Lilly, S.J., Hammer, F., Tresse, L., 1995, 455, 60.

Lemoine-Busserole, M., Contini, T., Pello, R., Le Borgner, J.-F., Kneib, J.-P., Lidman, C., 2003, astro-ph/0210547

MallÈn-Ornelas, G., Lilly, S.J., Crampton, D., Schade, D., 1999, ApJL, 518, 83.

McGaugh, S.S. 1991, ApJ, 380, 140.

Pagel, B.E.J., Edmunds, M.G., Blackwell, D.E., Chun, M.S., & Smith, G. 1979, MNRAS, 189, 95.

Pei, Y.C., Fall, S.M., & Hauser, M.G. 1999, ApJ, 522, 604.

Pettini, M., Shapley, A.E., Steidel, C.C., Cuby, J.G., Dickinson, M., Moorwood, A.F.M., Adelberger, K.L., & Giavalisco, M. 2001, ApJ, 554, 981.

Phillips, A. C., Guzm·n, R., Gallego, J., Koo, D. C., Lowenthal, J. D., Vogt, N. P., Faber, S. M., & Illingworth, G. D. 1997, ApJ, 489, 543

Prantzos, N. & Silk, J. 1998, ApJ, 507, 229.

Simard, L. & Pritchet, C. J. 1998, ApJ, 505, 96

Somerville, R. & Primack, J. 1999, MNRAS 310, 1087

Tresse, L., Maddox, S.J., Le Fevre, O., & Cuby, J.-G. 2002, MNRAS, 337, 369.

Twarog, B.A. 1980, ApJ, 242, 242

van Zee, L., Salzer, J.J., Haynes, M.P., O'Donoghue, A.A., & Balonek, T.J. 1998, AJ, 116, 2805.